%% file: main.tex
\documentclass[10pt,conference]{IEEEtran}

\usepackage{xspace}
\usepackage[table]{xcolor}
\usepackage{graphicx}
\usepackage[normalem]{ulem} %
\usepackage{amsmath}
\usepackage{amssymb}
\usepackage{amsfonts}
\usepackage{enumitem}
\usepackage{textcomp}

\usepackage{hyperref}
\hypersetup{linkcolor=black,citecolor=black,anchorcolor=black,filecolor=black,menucolor=black,runcolor=black,urlcolor=black,hidelinks}
\usepackage{breakurl}
\usepackage{cite}

\usepackage{booktabs}
\usepackage{multirow}
\usepackage{makecell}
\usepackage{ragged2e}

\usepackage{caption}
\usepackage{subcaption}

\usepackage{listings}

\usepackage{tikz}
\usetikzlibrary{calc}
\usetikzlibrary{shapes.geometric}
\usetikzlibrary{decorations.pathreplacing}
\usetikzlibrary{positioning}

\usepackage{flushend} %
\usepackage{datetime} %

\newcommand{\XSpace}[1]{}
\newcommand{\XComment}[1]{}
\newcommand{\Fix}[1]{\textcolor{red}{#1}}

\newcommand{\DefMacro}[2]{\expandafter\newcommand\csname rmk-#1\endcsname{#2}}
\newcommand{\UseMacro}[1]{\csname rmk-#1\endcsname}

\newcommand{\MyPara}[1]{\noindent\textbf{#1}.}
\newcommand{\MyParaOnly}[1]{\noindent\textbf{#1}}

\newcommand{\InputWithSpace}[1]{\bgroup\def\arraystretch{1.1}\input{#1}\egroup}
\newcommand{\Code}[1]{{\ifmmode{\mathtt{#1}}\else$\mathtt{#1}$\fi}}
\newcommand{\CodeIn}[1]{{\ifmmode{\mathtt{#1}}\else$\mathtt{#1}$\fi}}

\newcolumntype{R}[1]{>{\RaggedLeft\arraybackslash}p{#1}}
\newcolumntype{L}[1]{>{\RaggedRight\arraybackslash}p{#1}}

\definecolor{gray}{RGB}{211,211,211}
\newcommand{\jbasicstyle}{\small\sffamily} %

\newcommand{\jnumberstyle}{\scriptsize}

\lstdefinelanguage{pseudo}
{
morekeywords={},
keywordstyle=\bfseries,
lineskip=-0.1em,
numbers=left, %
numberstyle=\jnumberstyle,
numbersep=4pt,
basicstyle=\jbasicstyle,
breaklines=true,
breakautoindent=true,
tabsize=2,
columns=fullflexible,
morecomment=*[l][\textsl]{//},
mathescape=true,
xleftmargin=10pt,
}

\lstdefinelanguage{todo-comment}
{
morekeywords={},
keywordstyle=\bfseries,
lineskip=-0.1em,
numbers=none,
basicstyle=\jbasicstyle,
breaklines=true,
breakautoindent=true,
tabsize=2,
columns=fullflexible,
morecomment=*[l][\textsl]{//},
mathescape=true,
xleftmargin=-10pt,
}

\lstdefinelanguage{java-pretty}
{
language=java,
numbers=left,
basicstyle=\scriptsize\ttfamily,
numberstyle=\scriptsize,
breaklines=true,
columns=fullflexible,
xleftmargin=16pt,
showstringspaces=false,
}

\newcommand{\Tool}{\textsc{MoLE}\xspace}
\newcommand{\Title}{Mix-of-Language-Experts Architecture for Multilingual Programming}
\newcommand{\ToolURL}{https://github.com/uw-swag/mole}

\newcommand{\pl}{programming language\xspace}
\newcommand{\pls}{programming languages\xspace}
\newcommand{\lora}{LoRA\xspace}
\newcommand{\loras}{LoRAs\xspace}
\newcommand{\adapter}{adapter\xspace}
\newcommand{\adapters}{adapters\xspace}
\newcommand{\sharedadapter}{shared adapter\xspace}
\newcommand{\expertadapter}{expert adapter\xspace}
\newcommand{\expertadapters}{expert adapters\xspace}
\newcommand{\nladapter}{NL adapter\xspace}
\newcommand{\basemodel}{base model\xspace}
\newcommand{\codeblockticks}{\text{\textasciigrave\textasciigrave\textasciigrave}\xspace}
\newcommand{\describe}{summarization\xspace}
\newcommand{\Describe}{Summarization\xspace}
\newcommand{\synthesis}{synthesis\xspace}
\newcommand{\Synthesis}{Synthesis\xspace}
\newcommand{\translation}{translation\xspace}
\newcommand{\Translation}{Translation\xspace}
\newcommand{\peft}{parameter-efficient finetuning\xspace}

\newcommand{\glaive}{Glaive-code-assistant-v3\xspace}
\newcommand{\humanevalpack}{HumanEvalPack\xspace}
\newcommand{\humaneval}{HumanEval\xspace}
\newcommand{\deepseekcoder}{DeepSeek Coder\xspace}
\newcommand{\pissa}{PiSSA\xspace}

\DefMacro{model-pretrained}{No-FT\xspace}
\DefMacro{model-all-lang}{AllLang-FT\xspace}
\DefMacro{model-per-lang}{PerLang-FT\xspace}
\DefMacro{model-full-ft}{Full-FT\xspace}
\DefMacro{model-mole}{\Tool\xspace}
\DefMacro{model-mole-64-0}{\Tool-64+0\xspace}
\DefMacro{model-mole-56-8}{\Tool-56+8\xspace}
\DefMacro{model-mole-48-16}{\Tool-48+16\xspace}
\DefMacro{model-mole-32-32}{\Tool-32+32\xspace}
\DefMacro{model-mole-48-16-nl-expert}{NLExpert\xspace}
\DefMacro{model-mole-48-16-shared-last}{SharedLast\xspace}
\DefMacro{model-mole-48-16-std-init}{StdInit\xspace}

\DefMacro{lang-c}{C\xspace}
\DefMacro{lang-cpp}{C++\xspace}
\DefMacro{lang-csharp}{C\#\xspace}
\DefMacro{lang-go}{Go\xspace}
\DefMacro{lang-java}{Java\xspace}
\DefMacro{lang-js}{JS\xspace}
\DefMacro{lang-javascript}{\Fix{JS}\xspace}
\DefMacro{lang-python}{Python\xspace}
\DefMacro{lang-rust}{Rust\xspace}

\DefMacro{task-describe}{Summarization\xspace}
\DefMacro{task-synthesis}{Synthesis\xspace}
\DefMacro{task-translation}{Translation\xspace}

\DefMacro{metric-pass-1}{Pass@1\xspace}

\DefMacro{na}{N/A\xspace}
\DefMacro{avg}{Avg\xspace}

\DefMacro{TH-na}{N/A\xspace}
\DefMacro{TH-avg}{Avg\xspace}
\DefMacro{TH-model}{Model\xspace}
\DefMacro{TH-describe}{\Describe{}\xspace}
\DefMacro{TH-synthesis}{\Synthesis{}\xspace}
\DefMacro{TH-synthesize}{\Synthesis{}\xspace}
\DefMacro{TH-translation}{\Translation{}\xspace}

\input{tables/numbers-exp}
\input{tables/numbers-exp-manual}
\input{tables/numbers-exp-computed}

\begin{document}

\title{\Title}

\author{
\IEEEauthorblockN{Yifan Zong}
\IEEEauthorblockA{\textit{University of Waterloo}\\
Waterloo, Canada\\
y22zong@uwaterloo.ca}
\and
\IEEEauthorblockN{Yuntian Deng}
\IEEEauthorblockA{\textit{University of Waterloo}\\
Waterloo, Canada\\
yuntian@uwaterloo.ca}
\and
\IEEEauthorblockN{Pengyu Nie}
\IEEEauthorblockA{\textit{University of Waterloo}\\
Waterloo, Canada\\
pynie@uwaterloo.ca}
}

\maketitle

\begin{abstract}

Large language models (LLMs) have demonstrated impressive capabilities in aiding developers with tasks like code comprehension, generation, and translation.  Supporting multilingual programming---i.e., coding tasks across multiple \pls---typically requires either (1)~finetuning a single LLM across all \pls, which is cost-efficient but sacrifices language-specific specialization and performance, or (2)~finetuning separate LLMs for each \pl, which allows for specialization but is computationally expensive and storage-intensive due to the duplication of parameters.

This paper introduces \Tool (Mix-of-Language-Experts), a novel architecture that balances efficiency and specialization for multilingual programming.
\Tool is composed of a base model, a shared \lora (low-rank adaptation) module, and a collection of language-specific \lora modules.
These modules are jointly optimized during the finetuning process, enabling effective knowledge sharing and specialization across programming languages.
During inference, \Tool
automatically routes to the language-specific LoRA module corresponding to the programming language of the code token being generated.
Our experiments demonstrate that \Tool achieves greater parameter efficiency compared to training separate language-specific \loras, while outperforming a single shared LLM finetuned for all \pls in terms of accuracy.

\end{abstract}


\section{Introduction}
\label{sec:intro}

Large Language Models (LLMs) have transformed software development by enabling advanced code comprehension, generation, and translation capabilities~\cite{wang2023one,luo2023wizardcoder,shen2023pangu, WeiETAL24Magicoder, ChenETAL21Codex, GuoETAL24DeepSeekCoder}. Trained on massive code repositories, LLMs can enhance developer productivity by automating routine tasks and providing intelligent code completions. The success of tools like GitHub Copilot~\cite{Copilot}, Amazon CodeWhisperer~\cite{CodeWhisperer}, and Cursor~\cite{CursorAI} demonstrates the transformative impact of LLMs in software development.

To support multilingual programming---coding tasks across multiple \pls---current LLMs are typically pretrained and finetuned on diverse multilingual code repositories.
While this generalist approach allows broad applicability, it struggles with language-specific nuances, leading to suboptimal performance for individual languages.
To address this, language-specific finetuning has been explored, improving specialization but at significant computational and storage costs.
Moreover, independently finetuned models fail to share knowledge across languages, limiting performance in low-resource languages where data scarcity is common.

To bridge this gap, we propose \textbf{\Tool} (\textbf{M}ix-\textbf{o}f-\textbf{L}anguage-\textbf{E}xperts), a novel architecture that balances efficiency and specialization in multilingual programming.
\Tool employs a \emph{\sharedadapter} (used by programming language tokens) and an \emph{\nladapter} (used by natural language tokens) to store the common knowledge about programming patterns and concepts.
Another set of \emph{\expertadapters} store language-specific knowledge about each \pl's syntax and semantics.
During training, the language-agnostic and language-specific adapters are jointly optimized, enabling effective knowledge sharing while maintaining specialization.
At inference, an appropriate language-specific adapter is automatically activated according to each token's language.

Built on low-rank adaptation (\lora)~\cite{HuETAL21LoRA}, \Tool efficiently finetunes LLMs by updating only a small subset of parameters, with the frozen pretrained model serving as a foundation of multilingual programming knowledge.  Given that the pretrained model already contains multilingual programming knowledge in a less organized manner, the primary goal of \Tool's finetuning is to shift the language-agnostic and language-specific knowledge into the corresponding \lora \adapters.  To bootstrap this process, we apply a principal-components-based initialization strategy~\cite{MengETAL24Pissa} in \Tool, with the most important components assigned to the \sharedadapter, followed by the \expertadapters.

We evaluated \Tool on a multilingual code assistant dataset spanning eight programming languages and tested it across three tasks: code \describe, \synthesis, and \translation.  \Tool consistently outperforms both the all-language finetuning baseline (a single shared model for all languages) and per-language finetuning baseline (separate models for each language).  An analysis of individual expert performance confirms that \Tool effectively organizes language-agnostic and language-specific knowledge into distinct adapter components.

\noindent
The main contributions of this work include:

\begin{itemize}[topsep=0pt,itemsep=2pt,partopsep=0ex,parsep=0ex,leftmargin=*]

\item \MyPara{Architecture} A Mix-of-Language-Experts design that enables parameter-efficient finetuning for multilingual programming while separating language-agnostic and language-specific knowledge.%

\item \MyPara{Implementation} A fully open-source system that jointly finetunes adapters for eight \pls. %

\item \MyPara{Evaluation} Comprehensive experiments demonstrating \Tool's superior performance on code summarization, synthesis, and translation tasks. We also confirmed the effectiveness of our proposed architecture in disseminating knowledge across multiple \pls. %

\end{itemize}

\noindent
\Tool's code and model checkpoints are open-sourced at:\\
\url{\ToolURL}

\section{\Tool Technique}
\label{sec:technique}

\subsection{Architecture}
\label{sec:technique:arch}

\begin{figure}[t]
\centering
\includegraphics[width=.4\textwidth]{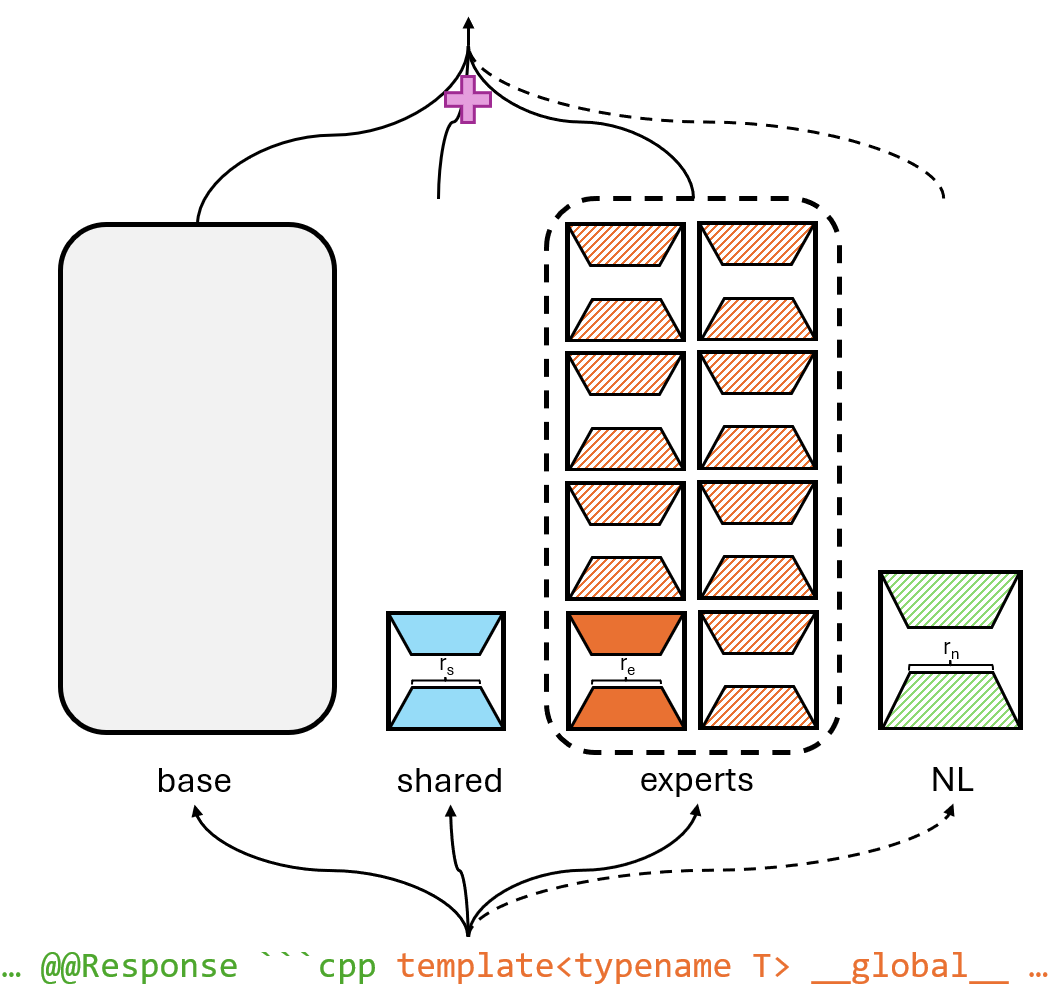}
\caption{Architecture of \Tool. The model extends the standard transformer by incorporating three types of LoRA adapters: a shared adapter (capturing commonalities across all programming languages), expert adapters (specializing in language-specific syntax and semantics), and an NL adapter (processing natural language tokens). During finetuning, the adapters are conditionally activated based on the input token's language, enabling efficient knowledge sharing and specialization.}
\label{fig:architecture}
\end{figure}

Figure~\ref{fig:architecture} illustrates \Tool's architecture, which extends the standard transformer architecture~\cite{vaswani2017attention} by incorporating three types of \lora \adapters~\cite{HuETAL21LoRA}: a \emph{\sharedadapter}, \emph{\expertadapters}, and a \emph{\nladapter}, in each linear layer of the feedforward network.  The original transformer model serves as the \emph{\basemodel}, while the adapters specialize in processing specific programming or natural languages, following the principle of separation of concerns~\cite{kulkarni2003separation}.

\lora~\cite{HuETAL21LoRA} is a parameter-efficient technique for finetuning transformer models under the assumption that parameter updates have a low intrinsic rank.  For a linear layer with weight matrix $W_0 \in \mathbb{R}^{d \times k}$, the \lora \adapter introduces two trainable matrices $B \in \mathbb{R}^{d \times r}$ and $A \in \mathbb{R}^{r \times k}$, where $r \ll min(d, k)$ is called the \emph{rank} of the adapter.  The weight update is then constrained to a low-rank decomposition $\Delta W = B A$, significantly reducing the number of trainable parameters.  For example, in a 1.3B parameter transformer model, a linear layer with $d \times k = 2048 \times 5504 \approx$ 11.2M parameters requires only 483k trainable parameters with $r = 64$ (4.3\% of the original layer's parameters).

\Tool leverages $K+2$ \lora \adapters for $K$ programming languages: a \sharedadapter ($\Delta W_s$), $K$ \expertadapters (denoted as $\Delta W_e$), and an \nladapter ($\Delta W_n$).  Adapters are conditionally activated based on the token's language:
\begin{itemize}
\item For \pl tokens:
$$
W = W_0 + \Delta W_s + \Delta W_e = W_0 + B_s A_s + B_e A_e
$$
\item For natural language tokens:
$$
W = W_0 + \Delta W_n = W_0 + B_n A_n
$$
\end{itemize}
All the \expertadapters have the same rank, and the rank of the \nladapter is the same as the combined ranks of the \sharedadapter and one \expertadapter, i.e., $r_n = r_s + r_e$.

The \sharedadapter captures common patterns across languages (e.g., control flow syntax), while \expertadapters specialize in language-specific constructs (e.g., Python's \CodeIn{elif} keyword).  The \nladapter processes natural language tokens, including user instructions and code explanations.

\subsection{Finetuning}
\label{sec:technique:finetuning}

\Tool finetunes the pretrained base transformer by disentangling its knowledge into adapters specialized for different languages. Unlike the ``noise \& zero'' initialization in standard \lora~\cite{HuETAL21LoRA}, %
\Tool uses a \emph{principal-components-based} initialization adapted from \pissa~\cite{MengETAL24Pissa}, leveraging the pretrained model's knowledge.  Specifically, for each linear layer with weight $W \in \mathbb{R}^{d \times k}$:

\begin{enumerate}[topsep=3pt,itemsep=1ex,partopsep=0ex,parsep=0ex,leftmargin=*]
\item Perform singular value decomposition (SVD):
$$
W = U S V^T.
$$
Here, $U \in \mathbb{R}^{d \times \min(d, k)}$ and $V \in \mathbb{R}^{k \times \min(d, k)}$ are singular vectors, and $S \in \mathbb{R}^{\min(d, k) \times \min(d, k)}$ is a diagonal matrix containing the singular values in descending order.

Conceptually, the larger singular values are, the more important the corresponding parameter components (represented by the singular vectors) are. Therefore, we assign the most important $r_n$ principal components to the \adapters in \Tool.

\item Initialize \lora adapters with assigned principle components:
Since the common knowledge is the basis of the language-specific knowledge, we assign the first $r_s$ principal components to the \sharedadapter, and the remaining $r_e$ principal components to the \expertadapters. The \nladapter, which is on an alternative computation flow, is always assigned all $r_n = r_s + r_e$ principal components.
The adapters are initialized as:
\begin{align*}
& A_s = U_{[:, :r_s]} S_{[:r_s, :r_s]}^{1/2}, B_s = S_{[:r_s, :r_s]}^{1/2} V_{[:, :r_s]}^T\\
& A_e = U_{[:, r_s:r_n]} S_{[r_s:r_n, r_s:r_n]}^{1/2}, B_e = S_{[r_s:r_n, r_s:r_n]}^{1/2} V_{[:, r_s:r_n]}^T\\
& A_n = U_{[:, :r_n]} S_{[:r_n, :r_n]}^{1/2}, B_n = S_{[:r_n, :r_n]}^{1/2} V_{[:, :r_n]}^T
\end{align*}
\item Finally, the residual principal components are used as the \basemodel's parameters:
$$
W_0 = U_{[:, r_n:]} S_{[r_n:, r_n:]} V_{[:, r_n:]}^T
$$
Our principal-components-based initialization strategy preserves the full capability of the pretrained model (i.e., $W = W_0 + B_s A_s + B_e A_e = W_0 + B_n A_n$), and allows each \adapter to evolve based on the \basemodel's knowledge during the finetuning process.
\end{enumerate}

\begin{figure}[t]
\centering
\includegraphics[width=.45\textwidth]{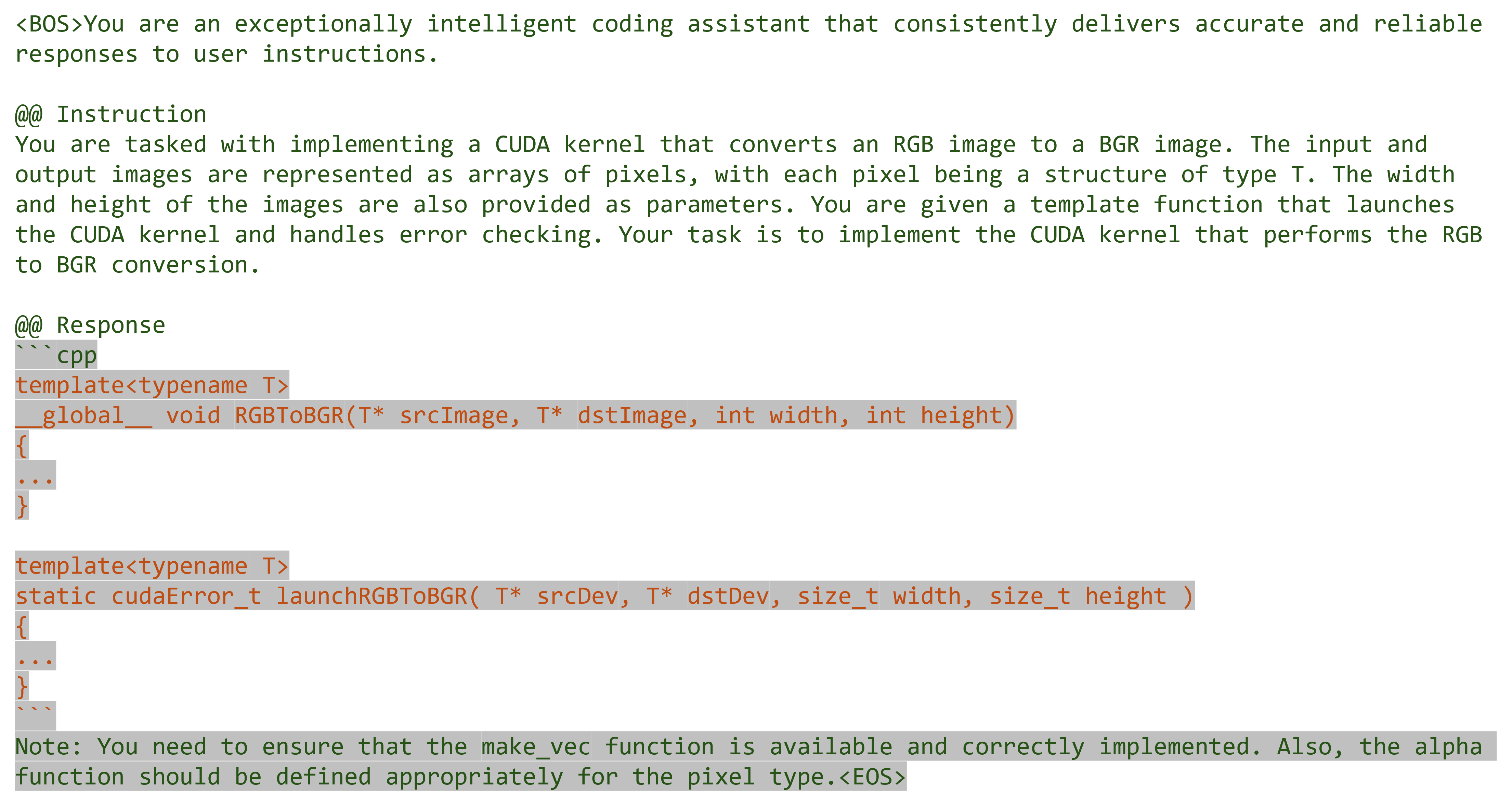}
\caption{Example finetuning data.  Green tokens represent natural language, while orange tokens represent \UseMacro{lang-cpp}. Grey background indicates the target tokens where loss is computed.}
\label{fig:example-finetuning}
\end{figure}

Finetuning is performed with the next-token prediction objective on the target tokens, activating \adapters based on token language labels. Figure~\ref{fig:example-finetuning} shows an example of the finetuning data, where the tokens belong to two languages: \UseMacro{lang-cpp} and natural language.  Involving multiple languages in the same finetuning data allows for the loss to back-propagate through \adapters that belong to different languages, and prompts the rearrangement of knowledge across different \adapters.

\subsection{Inference}
\label{sec:technique:inference}

During inference, \Tool dynamically activates adapters based on the token's language. Two modes are supported:

\begin{itemize}[topsep=3pt,itemsep=1ex,partopsep=0ex,parsep=0ex,leftmargin=*]
\item \MyPara{Automated Mode}
\Tool detects code block boundaries to switch adapters.
Specifically, \Tool starts off with using the \nladapter for generating natural language;
upon generating the code block opening tokens ``\CodeIn{\codeblockticks\{lang\}}'' (where \CodeIn{lang} is a \pl name, e.g., \CodeIn{python}), \Tool switches to use \CodeIn{lang}'s \expertadapter for generating code;
and then upon generating the code block closing token ``\CodeIn{\codeblockticks}'', \Tool switches back to the \nladapter for generating natural language.
This mode is useful for general purpose code assistant tasks, e.g., code generation (with optional natural language explanations) and code summarization.
For example, when generating the response in Figure~\ref{fig:example-finetuning}, \Tool will automatically switch from natural language to \UseMacro{lang-cpp} (for the code block) and back to natural language (for the notes in the end).
\item \MyPara{Manual Mode} The user specifies a target language, activating its corresponding adapter throughout generation.  This mode is useful when the output should be entirely in one specific \pl, e.g., code translation to a specific target language.  Figure~\ref{fig:example-translation} illustrates an example where \UseMacro{lang-cpp} is used exclusively for translation.
\end{itemize}

\section{Dataset}
\label{sec:data}

\subsection{Finetuning Dataset}
\label{sec:data:finetuning}

We choose the \glaive dataset~\cite{GlaiveDataset} as the finetuning dataset, because it covers a diverse set of \pls and is suitable for instruction-tuning general-purpose code assistants.
The dataset contains 950K coding problems and solutions generated using Glaive's synthetic data generation platform.
We classify dataset samples based on code block annotations and find that, although the dataset covers over 1,000 \pls, most have few data samples.   %
In this work, we focus on eight frequently-used \pls with the most data in the \glaive dataset: \UseMacro{lang-c}, \UseMacro{lang-cpp}, \UseMacro{lang-csharp}, \UseMacro{lang-go}, \UseMacro{lang-java}, \UseMacro{lang-js} (JavaScript), \UseMacro{lang-python}, and \UseMacro{lang-rust}.
We also observe that some of the problems and solutions are mostly in natural language (e.g., questions asking for third-party library suggestions), which diverge from the goal of this work to assist with programming tasks.
Thus, we set a filter that requires at least 1/3 of the characters in a sample (question + answer) belong to code blocks.
After filtering, we are left with a high-quality finetuning dataset of more than 192K samples and 7.86M tokens.
We randomly sample 5\% of data from each \pl as the validation set.

Table~\ref{tab:finetuning-dataset} shows the distribution of the number of samples over languages.
\UseMacro{lang-python} has the most amount of data, which may be due to its popularity in the training and evaluation sets of LLMs for code.
The number of samples in all languages are on a similar scale, with the lowest-resource language being \UseMacro{lang-go} with 9,533 samples.
An example of the finetuning data is shown in Figure~\ref{fig:example-finetuning}.

\begin{table}[t]
\centering
\caption{Statistics of the finetuning dataset.}
\label{tab:finetuning-dataset}

\begin{tabular}{l|r}
\toprule
\textbf{Language} & \textbf{\#Samples} \\
\midrule
\UseMacro{lang-c} & 15,773 \\
\UseMacro{lang-cpp} & 37,136 \\
\UseMacro{lang-csharp} & 11,320 \\
\UseMacro{lang-go} & 9,533 \\
\UseMacro{lang-java} & 24,010 \\
\UseMacro{lang-js} & 22,958 \\
\UseMacro{lang-python} & 52,642 \\
\UseMacro{lang-rust} & 18,835 \\
\midrule
\textbf{Total} & 192,207 \\
\bottomrule
\end{tabular}

\end{table}

\subsection{Evaluation Benchmarks}
\label{sec:data:benchmarks}

\begin{figure}[t]
\centering
\includegraphics[width=.45\textwidth]{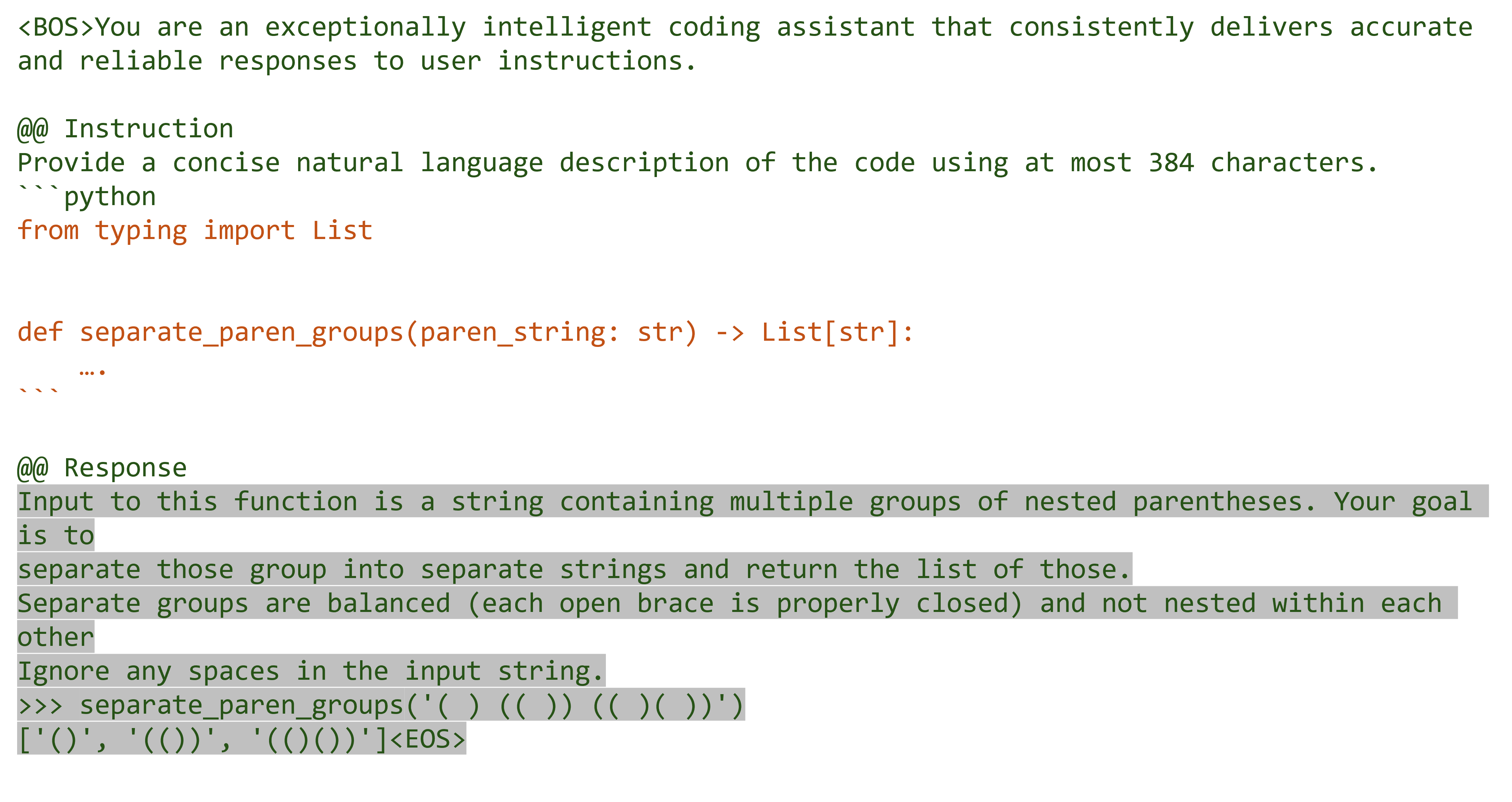}
\caption{Example of the \describe task. Green tokens represent natural language, while orange tokens represent \UseMacro{lang-python}. Tokens with a grey background indicate the target outputs that the model should generate.}
\label{fig:example-describe}
\end{figure}

\begin{figure}[t]
\centering
\includegraphics[width=.45\textwidth]{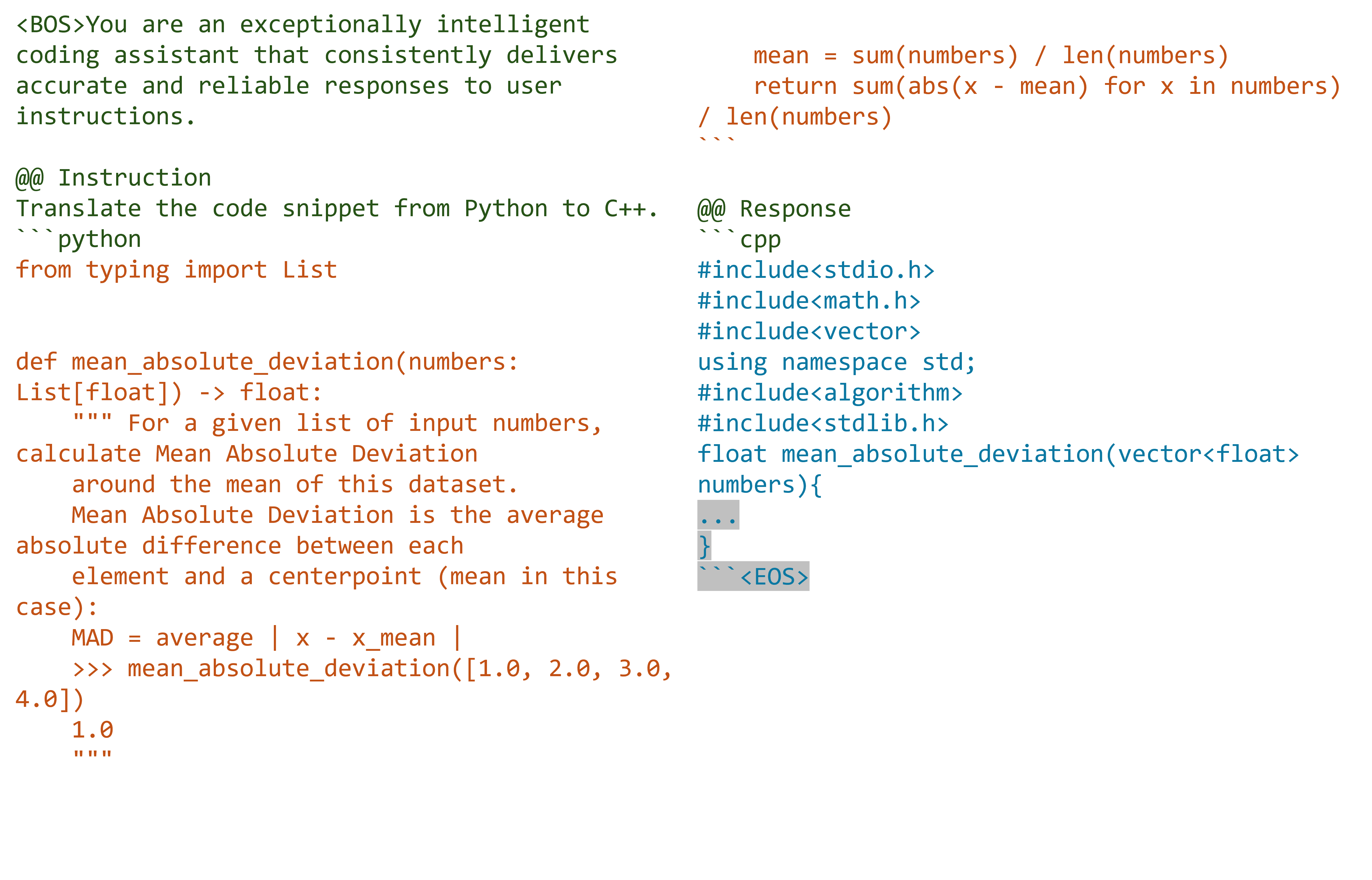}
\caption{Example of the \translation task. Green tokens represent natural language, orange tokens represent \UseMacro{lang-python}, and blue tokens correspond to \UseMacro{lang-cpp}. Tokens with a grey background indicate the target outputs that the model should generate.}
\label{fig:example-translation}
\end{figure}

We study the performance of \Tool on three distinct coding tasks: code \describe, \synthesis, and \translation.
To facilitate our evaluation, we build on the \humanevalpack benchmark~\cite{MuennighoffETAL24OctoPack,HumanEvalPackBenchmark}, which extends the Python-only \humaneval benchmark~\cite{ChenETAL21Codex} to six programming languages: \UseMacro{lang-cpp}, \UseMacro{lang-go}, \UseMacro{lang-java}, \UseMacro{lang-js}, \UseMacro{lang-python}, and \UseMacro{lang-rust}.

We directly adopt \humanevalpack's ``explain code'' and ``synthesize code'' tasks as the first two tasks in our evaluation, \describe and \synthesis.
Each task contains 164 samples per language, for a total of 984 samples.
Figure~\ref{fig:example-describe} shows an example of the \describe task, where the model is given a \UseMacro{lang-python} code snippet, and asked to generate a natural language summary of the code.

To study the performance of \Tool on the \translation task, we leverage the fact that \humanevalpack was developed by translating \humaneval's each Python sample to the other five languages.
Thus, we construct a \translation benchmark with all the 30 combinations of \{source language, target language\} among the six \pls in \humanevalpack, for a total of 4,920 samples.
Figure~\ref{fig:example-translation} shows an example of the \translation task, where the model is given a code snippet in \UseMacro{lang-python}, and asked to translate the code to \UseMacro{lang-cpp}.

\section{Evaluation}
\label{sec:eval}

We aim to address the following research questions:

\DefMacro{rq-vs-baselines}{RQ1}
\noindent\MyPara{\UseMacro{rq-vs-baselines}} How does \Tool perform compared to baseline \lora finetuning and full finetuning strategies?

\DefMacro{rq-ranks}{RQ2}
\noindent\MyPara{\UseMacro{rq-ranks}} What is the optimal distribution of shared adapter rank versus expert adapter rank in \Tool?

\DefMacro{rq-ablation}{RQ3}
\noindent\MyPara{\UseMacro{rq-ablation}} How do alternative design choices in \Tool, including the initialization approaches of \lora adapters and the usage of \nladapter, impact its performance?

\DefMacro{rq-experts}{RQ4}
\noindent\MyPara{\UseMacro{rq-experts}} Does each \expertadapter in \Tool effectively capture language-specific knowledge?

\subsection{Baselines}
\label{sec:eval:baseline}

We use \deepseekcoder 1.3B Base~\cite{GuoETAL24DeepSeekCoder,DeepSeekCoderModel} as the frozen pretrained model on which we finetune baselines and \Tool.
We consider the following baselines:

\MyParaOnly{\UseMacro{model-pretrained}} is the base pretrained model without any finetuning, used as the reference point to evaluate the benefit of finetuning.

\MyParaOnly{\UseMacro{model-all-lang}} is the model finetuned on all \pls, using \lora finetuning.  The \lora \adapter in this baseline has a rank of 64, to be comparable to \Tool.

\MyParaOnly{\UseMacro{model-per-lang}} is the combination of eight models, where each model is finetuned with the subset of the finetuning dataset that contains only one \pl.  Each model is \lora finetuned with rank of 64.  Upon inference, the model for the \pl corresponding to the problem's language will be used (this baseline is thus not applicable for the \translation task which involves two languages).

\MyParaOnly{\UseMacro{model-full-ft}} is the model finetuned on all \pls, with all parameters trainable. Note that \Tool, as a \peft technique, is \emph{not} designed to outperform full finetuning, so we use the full finetuning baseline as the ``upper-bound'' reference point when comparing \Tool and other \peft baselines.

\subsection{\Tool and Variants}
\label{sec:eval:tool}

We set the rank of the \nladapter (i.e., the sum of the \sharedadapter rank and \expertadapter rank) to 64.  We found that assigning rank 48 to the \sharedadapter and rank 16 to each \expertadapter achieves a good trade-off between learning common and specific \pl knowledge.  To study whether a larger or smaller \sharedadapter is beneficial, we experiment with the following distributions of \sharedadapter and \expertadapters:

\MyParaOnly{\UseMacro{model-mole-64-0}} uses \sharedadapter with rank 64 and no \expertadapter.  This variant learns only common \pl knowledge (but separeted from natural language knowledge which is handled by the \nladapter).

\MyParaOnly{\UseMacro{model-mole-56-8}} uses \sharedadapter with rank 56 and \expertadapters with rank 8.

\MyParaOnly{\UseMacro{model-mole-48-16}} (the default \Tool model) uses \sharedadapter with rank 48 and \expertadapters with rank 16.

\MyParaOnly{\UseMacro{model-mole-32-32}} uses \sharedadapter with rank 32 and \expertadapters with rank 32, to encourage the learning of language-specific knowledge.

\subsection{Metrics}
\label{sec:eval:metric}

Following \humanevalpack~\cite{MuennighoffETAL24OctoPack,HumanEvalPackBenchmark}, we use \UseMacro{metric-pass-1} consistently as the metric on \describe, \synthesis, and \translation tasks.
We use greedy decoding to generate one solution per problem.
For \synthesis and \translation tasks, the generated code solution is executed against the test cases that comes with the \humanevalpack benchmark to verify its correctness;
\UseMacro{metric-pass-1} is the percentage of problems that the generated solution passes all test cases.
For \describe task, the generated natural language summary is fed into the model again to synthesize a code snippet, whose correctness is verified by the test cases;
\UseMacro{metric-pass-1} is the percentage of problems where the code snippet synthesized from the generated natural language summary passes all test cases.

\subsection{Experiment Setup}
\label{sec:eval:setup}

\MyPara{Finetuning}
We finetune all models and baselines (except for \UseMacro{model-pretrained}) for two epochs using the AdamW optimizer.
We set the batch size to 64, and use a linear learning rate scheduler with 0.05 warmup.
The learning rate is set to 1e-4, which is based on a parameter search when finetuning \UseMacro{model-full-ft} and \UseMacro{model-all-lang} with learning rates of \{2e-5, 5e-5, 1e-4, 2e-4\}.
Each finetuning is repeated three times using different initial random seeds to account for the randomness of the training process, and we report the average results.

\MyPara{Hardware and Software Environment}
We conduct finetuning and evaluation on a GPU cluster with two types of GPUs: Nvidia A100 (40GB) and Nvidia Ada6000 (48GB); each finetuning or evaluation job is executed on one GPU.
We use Python 3.9, PyTorch 2.1.2, and Transformers 4.42.4.

\subsection{Results}
\label{sec:eval:result}

\begin{table}[t]
\centering
\caption{Summary of \UseMacro{metric-pass-1} results on all three tasks averaged over all \pls. The best \peft model (excluding \UseMacro{model-full-ft}) is highlighted.}
\label{tab:results-main}

\begin{tabular}{l | rrr}
\toprule
\textbf{\UseMacro{TH-model}} & \textbf{\UseMacro{TH-describe}} & \textbf{\UseMacro{TH-synthesis}} & \textbf{\UseMacro{TH-translation}} \\
\midrule
\UseMacro{model-pretrained}
&
\UseMacro{describe_pretrained_AVG}
&
\UseMacro{synthesize_pretrained_AVG}
&
\UseMacro{translation_pretrained_AVG}
\\
\UseMacro{model-all-lang}
&
\UseMacro{describe_all-lang_AVG}
&
\UseMacro{synthesize_all-lang_AVG}
&
\UseMacro{translation_all-lang_AVG}
\\
\UseMacro{model-per-lang}
&
\UseMacro{describe_per-lang_AVG}
&
\UseMacro{synthesize_per-lang_AVG}
&
\UseMacro{TH-na}
\\
{\color{gray!50!black} \UseMacro{model-full-ft}}
&
{\color{gray!50!black} \UseMacro{describe_full-ft_AVG}}
&
{\color{gray!50!black} \UseMacro{synthesize_full-ft_AVG}}
&
{\color{gray!50!black} \UseMacro{translation_full-ft_AVG}}
\\
\midrule
\UseMacro{model-mole}
&
\textbf{\UseMacro{describe_mole-48-16_AVG}}
&
\textbf{\UseMacro{synthesize_mole-48-16_AVG}}
&
\textbf{\UseMacro{translation_mole-48-16_AVG}}
\\
\bottomrule
\end{tabular}

\end{table}

Table~\ref{tab:results-main} presents the summary of results on all three tasks of code \describe, \synthesis, and \translation, where the numbers are the average \UseMacro{metric-pass-1} over all data samples across all \pls on the corresponding task.

We found that \Tool outperforms both \UseMacro{model-all-lang} and \UseMacro{model-per-lang} on all three tasks, confirming \Tool's ability in assisting multilingual programming.
As expected, finetuning one model per \pl (i.e., \UseMacro{model-per-lang}) outperforms sharing a single model for all \pls (i.e., \UseMacro{model-all-lang}), but is only marginally better.
Compared to \UseMacro{model-per-lang} which finetunes each \lora \adapter on a smaller subset, the \expertadapters in \Tool are jointly trained, and thus can achieve a better organization of specific \pl knowledge, and benefit from the \sharedadapter and \nladapter for processing common \pl and natural language knowledge.
Notably, \Tool's performance is on a par with the full finetuning baseline (\UseMacro{model-full-ft}), even outperforming it on the \describe task, which shows the effectiveness of \Tool's architecture under the \peft paradigm.

\begin{table}[t]
\begin{scriptsize}
\centering
\caption{\UseMacro{metric-pass-1} results on the \describe task for each \pl. The best \peft model (excluding \UseMacro{model-full-ft}) is highlighted.}
\label{tab:results-describe}
\vspace{-2pt}

\begin{tabular}{l | rrrrrr | r}
\toprule
\textbf{\UseMacro{TH-model}}
& \textbf{\UseMacro{lang-cpp}}
& \textbf{\UseMacro{lang-go}}
& \textbf{\UseMacro{lang-java}}
& \textbf{\UseMacro{lang-js}}
& \textbf{\UseMacro{lang-python}}
& \textbf{\UseMacro{lang-rust}}
& \textbf{\UseMacro{TH-avg}} \\
\midrule
\UseMacro{model-pretrained}
&
\UseMacro{describe_pretrained_cpp}
&
\UseMacro{describe_pretrained_go}
&
\UseMacro{describe_pretrained_java}
&
\UseMacro{describe_pretrained_js}
&
\UseMacro{describe_pretrained_python}
&
\UseMacro{describe_pretrained_rust}
&
\UseMacro{describe_pretrained_AVG}
\\
\UseMacro{model-all-lang}
&
\textbf{\UseMacro{describe_all-lang_cpp}}
&
\UseMacro{describe_all-lang_go}
&
\UseMacro{describe_all-lang_java}
&
\UseMacro{describe_all-lang_js}
&
\UseMacro{describe_all-lang_python}
&
\UseMacro{describe_all-lang_rust}
&
\UseMacro{describe_all-lang_AVG}
\\
\UseMacro{model-per-lang}
&
\UseMacro{describe_per-lang_cpp}
&
\UseMacro{describe_per-lang_go}
&
\UseMacro{describe_per-lang_java}
&
\UseMacro{describe_per-lang_js}
&
\textbf{\UseMacro{describe_per-lang_python}}
&
\UseMacro{describe_per-lang_rust}
&
\UseMacro{describe_per-lang_AVG}
\\
{\color{gray!50!black} \UseMacro{model-full-ft}}
&
{\color{gray!50!black} \UseMacro{describe_full-ft_cpp}}
&
{\color{gray!50!black} \UseMacro{describe_full-ft_go}}
&
{\color{gray!50!black} \UseMacro{describe_full-ft_java}}
&
{\color{gray!50!black} \UseMacro{describe_full-ft_js}}
&
{\color{gray!50!black} \UseMacro{describe_full-ft_python}}
&
{\color{gray!50!black} \UseMacro{describe_full-ft_rust}}
&
{\color{gray!50!black} \UseMacro{describe_full-ft_AVG}}
\\
\midrule
\UseMacro{model-mole-64-0}
&
\UseMacro{describe_mole-64-0_cpp}
&
\UseMacro{describe_mole-64-0_go}
&
\UseMacro{describe_mole-64-0_java}
&
\UseMacro{describe_mole-64-0_js}
&
\UseMacro{describe_mole-64-0_python}
&
\UseMacro{describe_mole-64-0_rust}
&
\UseMacro{describe_mole-64-0_AVG}
\\
\UseMacro{model-mole-56-8}
&
\UseMacro{describe_mole-56-8_cpp}
&
\UseMacro{describe_mole-56-8_go}
&
\UseMacro{describe_mole-56-8_java}
&
\UseMacro{describe_mole-56-8_js}
&
\UseMacro{describe_mole-56-8_python}
&
\UseMacro{describe_mole-56-8_rust}
&
\UseMacro{describe_mole-56-8_AVG}
\\
\UseMacro{model-mole-48-16}
&
\UseMacro{describe_mole-48-16_cpp}
&
\UseMacro{describe_mole-48-16_go}
&
\textbf{\UseMacro{describe_mole-48-16_java}}
&
\UseMacro{describe_mole-48-16_js}
&
\textbf{\UseMacro{describe_mole-48-16_python}}
&
\textbf{\UseMacro{describe_mole-48-16_rust}}
&
\textbf{\UseMacro{describe_mole-48-16_AVG}}
\\
\UseMacro{model-mole-32-32}
&
\UseMacro{describe_mole-32-32_cpp}
&
\textbf{\UseMacro{describe_mole-32-32_go}}
&
\UseMacro{describe_mole-32-32_java}
&
\textbf{\UseMacro{describe_mole-32-32_js}}
&
\UseMacro{describe_mole-32-32_python}
&
\UseMacro{describe_mole-32-32_rust}
&
\UseMacro{describe_mole-32-32_AVG}
\\
\bottomrule
\end{tabular}

\vspace{-5pt}
\end{scriptsize}
\end{table}

\begin{table}[t]
\begin{scriptsize}
\centering
\caption{\UseMacro{metric-pass-1} results on the \synthesis task for each \pl. The best \peft model (excluding \UseMacro{model-full-ft}) is highlighted.}
\label{tab:results-synthesize}
\vspace{-2pt}

\begin{tabular}{l | rrrrrr | r}
\toprule
\textbf{\UseMacro{TH-model}}
& \textbf{\UseMacro{lang-cpp}}
& \textbf{\UseMacro{lang-go}}
& \textbf{\UseMacro{lang-java}}
& \textbf{\UseMacro{lang-js}}
& \textbf{\UseMacro{lang-python}}
& \textbf{\UseMacro{lang-rust}}
& \textbf{\UseMacro{TH-avg}} \\
\midrule
\UseMacro{model-pretrained}
&
\textbf{\UseMacro{synthesize_pretrained_cpp}}
&
\UseMacro{synthesize_pretrained_go}
&
\UseMacro{synthesize_pretrained_java}
&
\UseMacro{synthesize_pretrained_js}
&
\UseMacro{synthesize_pretrained_python}
&
\UseMacro{synthesize_pretrained_rust}
&
\UseMacro{synthesize_pretrained_AVG}
\\
\UseMacro{model-all-lang}
&
\UseMacro{synthesize_all-lang_cpp}
&
\UseMacro{synthesize_all-lang_go}
&
\UseMacro{synthesize_all-lang_java}
&
\UseMacro{synthesize_all-lang_js}
&
\UseMacro{synthesize_all-lang_python}
&
\UseMacro{synthesize_all-lang_rust}
&
\UseMacro{synthesize_all-lang_AVG}
\\
\UseMacro{model-per-lang}
&
\UseMacro{synthesize_per-lang_cpp}
&
\UseMacro{synthesize_per-lang_go}
&
\textbf{\UseMacro{synthesize_per-lang_java}}
&
\UseMacro{synthesize_per-lang_js}
&
\UseMacro{synthesize_per-lang_python}
&
\textbf{\UseMacro{synthesize_per-lang_rust}}
&
\UseMacro{synthesize_per-lang_AVG}
\\
{\color{gray!50!black} \UseMacro{model-full-ft}}
&
{\color{gray!50!black} \UseMacro{synthesize_full-ft_cpp}}
&
{\color{gray!50!black} \UseMacro{synthesize_full-ft_go}}
&
{\color{gray!50!black} \UseMacro{synthesize_full-ft_java}}
&
{\color{gray!50!black} \UseMacro{synthesize_full-ft_js}}
&
{\color{gray!50!black} \UseMacro{synthesize_full-ft_python}}
&
{\color{gray!50!black} \UseMacro{synthesize_full-ft_rust}}
&
{\color{gray!50!black} \UseMacro{synthesize_full-ft_AVG}}
\\
\midrule
\UseMacro{model-mole-64-0}
&
\UseMacro{synthesize_mole-64-0_cpp}
&
\UseMacro{synthesize_mole-64-0_go}
&
\UseMacro{synthesize_mole-64-0_java}
&
\UseMacro{synthesize_mole-64-0_js}
&
\UseMacro{synthesize_mole-64-0_python}
&
\UseMacro{synthesize_mole-64-0_rust}
&
\UseMacro{synthesize_mole-64-0_AVG}
\\
\UseMacro{model-mole-56-8}
&
\UseMacro{synthesize_mole-56-8_cpp}
&
\textbf{\UseMacro{synthesize_mole-56-8_go}}
&
\UseMacro{synthesize_mole-56-8_java}
&
\UseMacro{synthesize_mole-56-8_js}
&
\UseMacro{synthesize_mole-56-8_python}
&
\UseMacro{synthesize_mole-56-8_rust}
&
\UseMacro{synthesize_mole-56-8_AVG}
\\
\UseMacro{model-mole-48-16}
&
\UseMacro{synthesize_mole-48-16_cpp}
&
\textbf{\UseMacro{synthesize_mole-48-16_go}}
&
\UseMacro{synthesize_mole-48-16_java}
&
\UseMacro{synthesize_mole-48-16_js}
&
\textbf{\UseMacro{synthesize_mole-48-16_python}}
&
\UseMacro{synthesize_mole-48-16_rust}
&
\UseMacro{synthesize_mole-48-16_AVG}
\\
\UseMacro{model-mole-32-32}
&
\UseMacro{synthesize_mole-32-32_cpp}
&
\UseMacro{synthesize_mole-32-32_go}
&
\UseMacro{synthesize_mole-32-32_java}
&
\textbf{\UseMacro{synthesize_mole-32-32_js}}
&
\UseMacro{synthesize_mole-32-32_python}
&
\UseMacro{synthesize_mole-32-32_rust}
&
\textbf{\UseMacro{synthesize_mole-32-32_AVG}}
\\
\bottomrule
\end{tabular}

\vspace{-5pt}
\end{scriptsize}
\end{table}

\begin{figure*}[t]
\centering
\includegraphics[width=.85\textwidth]{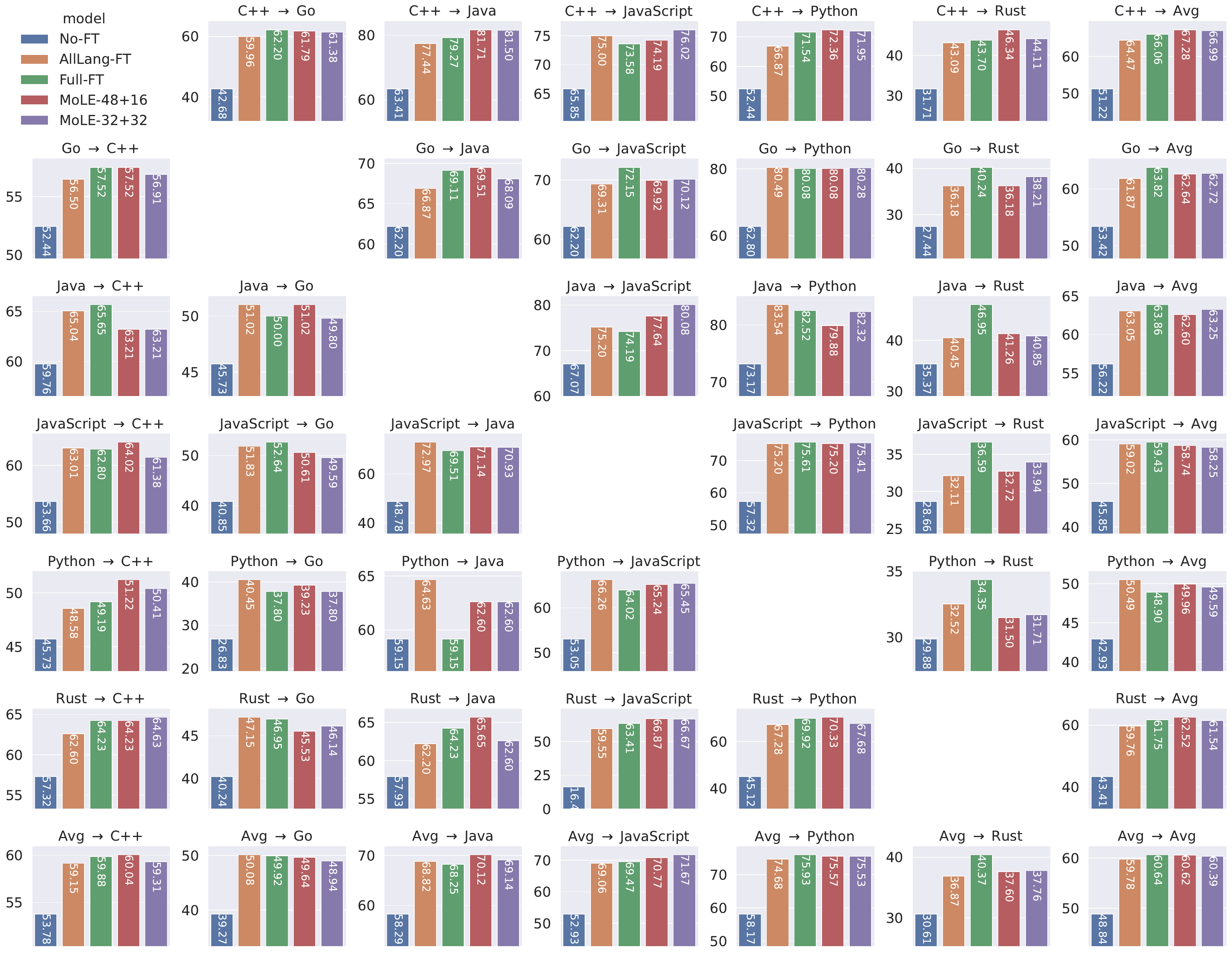}
\caption{Barplots showing \UseMacro{metric-pass-1} on the \translation task for each source-target \pl pair. The last column shows the average \UseMacro{metric-pass-1} when translating \emph{from} a \pl to others; the last row shows the average \UseMacro{metric-pass-1} when translating from others \emph{to} a \pl.}
\label{fig:results-translation}
\end{figure*}

Tables~\ref{tab:results-describe} and \ref{tab:results-synthesize} present the results per \pl for the \describe and \synthesis tasks, respectively.
We found that \Tool outperforms the baselines in most cases, especially on low-resource languages such as \UseMacro{lang-go} and \UseMacro{lang-js}.
For example, on the \describe task, \UseMacro{metric-pass-1} of \UseMacro{lang-js} is improved from \UseMacro{describe_per-lang_js}\% (of \UseMacro{model-per-lang}) up to \UseMacro{describe_mole-32-32_js}\% (of \UseMacro{model-mole-32-32});
on the \synthesis task, \UseMacro{metric-pass-1} of \UseMacro{lang-js} is improved from \UseMacro{synthesize_per-lang_js}\% (of \UseMacro{model-per-lang}) up to \UseMacro{synthesize_mole-32-32_js}\% (of \UseMacro{model-mole-32-32}).
For \UseMacro{lang-python}, the performance of \UseMacro{model-mole} is the same as or slightly better than that of baselines, which can be attributed to the fact that \UseMacro{lang-python} is prevalent in the pretraining dataset of the base model.
We observe minor performance drops for \UseMacro{lang-cpp}, \UseMacro{lang-java} on \synthesis task, and \UseMacro{lang-rust} on \describe task when using \Tool; we hypothesize that these ``harder'' \pls require stronger language-specific knowledge to excel, and our finetuning data may be insufficient to train good \expertadapters for them.
Nevertheless, \Tool still achieves an overall improvement over the baselines when considering the average of all \pls, making it a suitable choice for multilingual programming.

\figurename~\ref{fig:results-translation} presents the results on the \translation task for each pair of source and target \pl.
We observe a similar trend as in the \describe and \synthesis tasks, where \Tool achieves the most improvements over the baselines on low-resource languages.
For example, when translating from other \pls to \UseMacro{lang-rust}, \Tool achieves an average \UseMacro{metric-pass-1} of \UseMacro{translation_mole-32-32_AVG_rust}\%, outperforming the \UseMacro{model-all-lang} baseline's \UseMacro{translation_all-lang_AVG_rust}\%;
when translating from \UseMacro{lang-rust} to other \pls, the improvement is even larger, with \Tool's \UseMacro{translation_mole-48-16_rust_AVG}\% vs. \UseMacro{model-all-lang}'s \UseMacro{translation_all-lang_rust_AVG}\%.

\MyPara{Answer to \UseMacro{rq-vs-baselines}} \Tool outperforms both \UseMacro{model-all-lang} and \UseMacro{model-per-lang} baselines on code \describe, \synthesis, and \translation tasks.
This indicates that \Tool's architecture allows more efficient finetuning of multilingual programming models.
In particular, \Tool can greatly benefit low-resource languages by jointly training with other popular languages.

Next, we study the impact of the rank distribution between the \sharedadapter and \expertadapters on the performance of \Tool.
The lower parts of Tables~\ref{tab:results-describe} and \ref{tab:results-synthesize} present the \UseMacro{metric-pass-1} of \Tool variants with different rank distributions.
We found that not using \expertadapters at all (i.e., \UseMacro{model-mole-64-0}) leads to worse performance, indicating that learning specific \pl knowledge is important in \Tool's architecture.
The other three variants are have comparable performance;
\UseMacro{model-mole-48-16} is the best on the \describe task, and \UseMacro{model-mole-32-32} is more effective on the \synthesis task.
On the \translation task, we compare the performance of \UseMacro{model-mole-48-16} and \UseMacro{model-mole-32-32} variants, as shown by the last two bars in the barplots in Figure~\ref{fig:results-translation}.
\UseMacro{model-mole-48-16} outperforms \UseMacro{model-mole-32-32} by a small margin, with same or better performance on 18 out of 30 source-target language pairs.

\MyPara{Answer to \UseMacro{rq-ranks}}
The use of \expertadapters is critical in improving \Tool's multilingual programming ability.
Overall, the \UseMacro{model-mole-48-16} variant achieves the best performance on two out of the three tasks (\describe and \translation) and is comparable to the best performing variant on the \synthesis task (\UseMacro{model-mole-32-32}), thus we recommend \UseMacro{model-mole-48-16} as the default \Tool configuration.

\begin{table}[t]
\begin{scriptsize}
\centering
\caption{Ablation study on the design choices of \Tool: \UseMacro{model-mole-48-16-std-init} uses the standard \lora initialization instead of \pissa; \UseMacro{model-mole-48-16-shared-last} assigns the most significant components to the \expertadapters instead of the \sharedadapter; \UseMacro{model-mole-48-16-nl-expert} processes natural language using an \expertadapter.}
\label{tab:results-ablation}
\begin{minipage}{0.48\textwidth}
\centering
\subcaption{\UseMacro{metric-pass-1} on the \describe task.}
\label{tab:results-ablation-describe}

\begin{tabular}{l | rrrrrr | r}
\toprule
\textbf{\UseMacro{TH-model}}
& \textbf{\UseMacro{lang-cpp}}
& \textbf{\UseMacro{lang-go}}
& \textbf{\UseMacro{lang-java}}
& \textbf{\UseMacro{lang-js}}
& \textbf{\UseMacro{lang-python}}
& \textbf{\UseMacro{lang-rust}}
& \textbf{\UseMacro{TH-avg}} \\
\midrule
\UseMacro{model-mole}
&
\UseMacro{describe_mole-48-16_cpp}
&
\UseMacro{describe_mole-48-16_go}
&
\UseMacro{describe_mole-48-16_java}
&
\UseMacro{describe_mole-48-16_js}
&
\UseMacro{describe_mole-48-16_python}
&
\UseMacro{describe_mole-48-16_rust}
&
\UseMacro{describe_mole-48-16_AVG}
\\
\midrule
\UseMacro{model-mole-48-16-std-init}
&
\UseMacro{describe_mole-48-16-std-init_cpp}
&
\UseMacro{describe_mole-48-16-std-init_go}
&
\UseMacro{describe_mole-48-16-std-init_java}
&
\UseMacro{describe_mole-48-16-std-init_js}
&
\UseMacro{describe_mole-48-16-std-init_python}
&
\UseMacro{describe_mole-48-16-std-init_rust}
&
\UseMacro{describe_mole-48-16-std-init_AVG}
\\
\UseMacro{model-mole-48-16-shared-last}
&
\UseMacro{describe_mole-48-16-shared-last_cpp}
&
\UseMacro{describe_mole-48-16-shared-last_go}
&
\UseMacro{describe_mole-48-16-shared-last_java}
&
\UseMacro{describe_mole-48-16-shared-last_js}
&
\UseMacro{describe_mole-48-16-shared-last_python}
&
\UseMacro{describe_mole-48-16-shared-last_rust}
&
\UseMacro{describe_mole-48-16-shared-last_AVG}
\\
\UseMacro{model-mole-48-16-nl-expert}
&
\UseMacro{describe_mole-48-16-nl-expert_cpp}
&
\UseMacro{describe_mole-48-16-nl-expert_go}
&
\UseMacro{describe_mole-48-16-nl-expert_java}
&
\UseMacro{describe_mole-48-16-nl-expert_js}
&
\UseMacro{describe_mole-48-16-nl-expert_python}
&
\UseMacro{describe_mole-48-16-nl-expert_rust}
&
\UseMacro{describe_mole-48-16-nl-expert_AVG}
\\
\bottomrule
\end{tabular}

\end{minipage}
\hfill
\vspace{5pt}
\begin{minipage}{0.48\textwidth}
\centering
\subcaption{\UseMacro{metric-pass-1} on the \synthesis task.}
\label{tab:results-ablation-synthesize}

\begin{tabular}{l | rrrrrr | r}
\toprule
\textbf{\UseMacro{TH-model}}
& \textbf{\UseMacro{lang-cpp}}
& \textbf{\UseMacro{lang-go}}
& \textbf{\UseMacro{lang-java}}
& \textbf{\UseMacro{lang-js}}
& \textbf{\UseMacro{lang-python}}
& \textbf{\UseMacro{lang-rust}}
& \textbf{\UseMacro{TH-avg}} \\
\midrule
\UseMacro{model-mole}
&
\UseMacro{synthesize_mole-48-16_cpp}
&
\UseMacro{synthesize_mole-48-16_go}
&
\UseMacro{synthesize_mole-48-16_java}
&
\UseMacro{synthesize_mole-48-16_js}
&
\UseMacro{synthesize_mole-48-16_python}
&
\UseMacro{synthesize_mole-48-16_rust}
&
\UseMacro{synthesize_mole-48-16_AVG}
\\
\midrule
\UseMacro{model-mole-48-16-std-init}
&
\UseMacro{synthesize_mole-48-16-std-init_cpp}
&
\UseMacro{synthesize_mole-48-16-std-init_go}
&
\UseMacro{synthesize_mole-48-16-std-init_java}
&
\UseMacro{synthesize_mole-48-16-std-init_js}
&
\UseMacro{synthesize_mole-48-16-std-init_python}
&
\UseMacro{synthesize_mole-48-16-std-init_rust}
&
\UseMacro{synthesize_mole-48-16-std-init_AVG}
\\
\UseMacro{model-mole-48-16-shared-last}
&
\UseMacro{synthesize_mole-48-16-shared-last_cpp}
&
\UseMacro{synthesize_mole-48-16-shared-last_go}
&
\UseMacro{synthesize_mole-48-16-shared-last_java}
&
\UseMacro{synthesize_mole-48-16-shared-last_js}
&
\UseMacro{synthesize_mole-48-16-shared-last_python}
&
\UseMacro{synthesize_mole-48-16-shared-last_rust}
&
\UseMacro{synthesize_mole-48-16-shared-last_AVG}
\\
\UseMacro{model-mole-48-16-nl-expert}
&
\UseMacro{synthesize_mole-48-16-nl-expert_cpp}
&
\UseMacro{synthesize_mole-48-16-nl-expert_go}
&
\UseMacro{synthesize_mole-48-16-nl-expert_java}
&
\UseMacro{synthesize_mole-48-16-nl-expert_js}
&
\UseMacro{synthesize_mole-48-16-nl-expert_python}
&
\UseMacro{synthesize_mole-48-16-nl-expert_rust}
&
\UseMacro{synthesize_mole-48-16-nl-expert_AVG}
\\
\bottomrule
\end{tabular}

\end{minipage}
\end{scriptsize}
\end{table}

We perform an ablation study to investigate the impact of different design choices in \Tool, the results of which are presented in Table~\ref{tab:results-ablation}.
\Tool adapted the \pissa~\cite{MengETAL24Pissa} initialization technique for the \lora \adapters, which promotes a better organization of language-agnostic and language-specific knowledge and speeds up the convergence of the finetuning process.
Compared with the standard \lora initialization (\UseMacro{model-mole-48-16-std-init}), the final performance is close, favoring the \Tool on the \describe task and the \UseMacro{lang-go} and \UseMacro{lang-python} \pls.
\Tool also positions the \sharedadapter before the \expertadapters in the principle components' space, with the hypothesis that the language-agnostic knowledge is more important than the language-specific knowledge.
The \UseMacro{model-mole-48-16-shared-last} variant explores the reversed order by positioning the \sharedadapter after the \expertadapters.
Results show that \Tool consistently outperforms \UseMacro{model-mole-48-16-shared-last} on both tasks, confirming our hypothesis.
Finally, \Tool treats natural language specially by using a separate rank-64 \nladapter, since the natural language knowledge should be separated from \pl knowledge.
To verify this intuition, we examine \UseMacro{model-mole-48-16-nl-expert}, which processes natural language with an additional \expertadapter instead of a separate \nladapter (i.e. treat natural language as one of the \pls).
Overall, \Tool outperforms \UseMacro{model-mole-48-16-nl-expert}, with larger gap on the \describe task than the \synthesis task, indicating that it is more beneficial to learn natural language knowledge in a separate adapter.

\MyPara{Answer to \UseMacro{rq-ablation}} In \Tool's architecture, the order of the \sharedadapter and \expertadapters in the principle components' space is critical, and a separate \nladapter is essential for processing natural language knowledge.

\begin{figure*}[t]
\centering
\includegraphics[width=.9\textwidth]{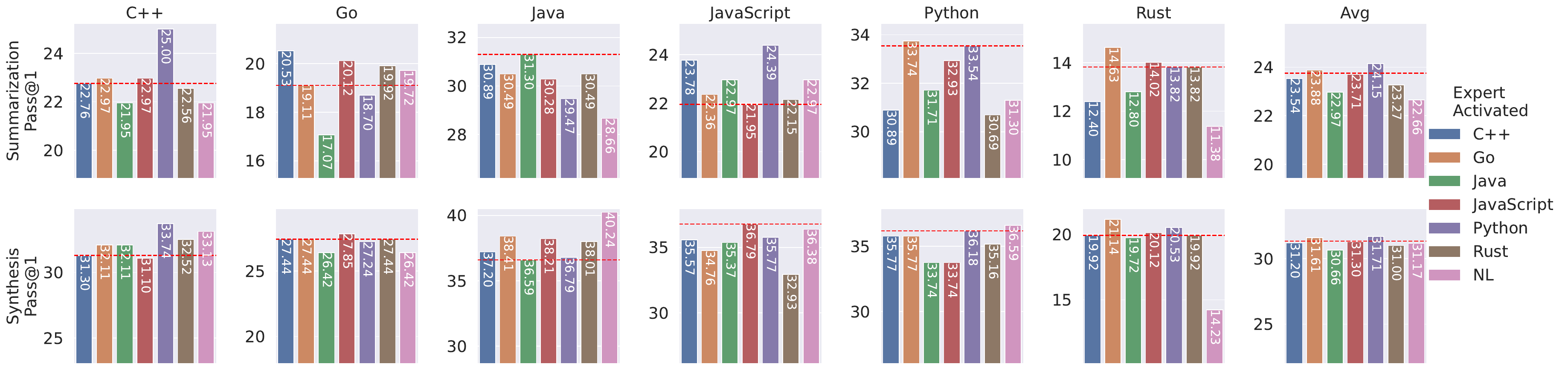}
\caption{Performance of \Tool when forcing the model to activate one \expertadapter or \nladapter in the \describe (top) and \synthesis (bottom) tasks. The red dotted line in each subplot represents the performance of the \Tool variant when activating the correct adapter matching the input/output.}
\label{fig:mismatch}
\end{figure*}

We further investigate whether each \expertadapter in \Tool is properly learning its assigned \pl.
To quantitatively measure this, we manually activate a misassigned adapter during inference (using the same finetuned \Tool model), and compare the performance with when the correct adapter is activated.
The results are shown in Figure~\ref{fig:mismatch}.
On average (last column in Figure~\ref{fig:mismatch}), activating the incorrect \expertadapter leads to worse performance in most of the cases, with the exception of the \expertadapter for \UseMacro{lang-go} and \UseMacro{lang-python}.
We suspect that these two \expertadapters are stronger than others for different reasons:
\UseMacro{lang-go} is syntactically close to several other \pls being studied (e.g., \UseMacro{lang-cpp}, \UseMacro{lang-java}, and \UseMacro{lang-rust}), thus using it to summarize or generate code for other \pls achieves moderate performance;
the \UseMacro{lang-python} \expertadapter is the most well-trained one, given that \UseMacro{lang-python} data takes the largest proportion in both the finetuning dataset of \Tool and the pretraining dataset of the base model.
Looking at the detailed results on each \pl, we found that applying the \expertadapter of one \pl to a similar \pl's input/output will likely achieve good performance, e.g., using \UseMacro{lang-cpp} \expertadapter on \UseMacro{lang-go} and verse versa both achieve similar or higher performance than using the correct \expertadapter.

We also experimented with using only the \nladapter throughout the inference, i.e., considering all tokens as natural language, as shown by the ``NL'' bar in Figure~\ref{fig:mismatch}.
Except for a few outliers (e.g., \UseMacro{lang-java} on the \synthesis task), the performance of forcing the \nladapter is lower than using the correct adapter, indicating that the natural language knowledge and \pl knowledge are also properly disentangled in \Tool.

\MyPara{Answer to \UseMacro{rq-experts}} Our analysis on the \UseMacro{metric-pass-1} when activating \Tool's \expertadapters on mismatching languages confirms that each \expertadapter is learning language-specific knowledge as expected.  In some cases, the \expertadapter for syntactically similar \pls (e.g., \UseMacro{lang-cpp} and \UseMacro{lang-go}) can be exchanged without significant performance drop.

\section{Limitations and Future Work}
\label{sec:discussion}

\Tool is the first architecture designed for multilingual programming with explicit expert \adapter modules.
Constrained by the accessible computational resources, our implementation is based on an 1.3B LLM and uses \lora for \peft.
The same idea can be experimented on larger LLMs, extended to full finetuning, or even applied in the pretraining phase.
We believe that the proposed shared-and-expert \adapters architecture will lead to better multilingual programming performance on larger-scale LLMs.

The current finetuning dataset of \Tool is limited in its scale and modality.
Our results analysis shows that some of \Tool's \expertadapters are not sufficiently trained compared to the \pls with sufficient data (e.g., \UseMacro{lang-python}).
Combining similar languages into a single \expertadapter can further improve parameter efficiency while alleviating the under-training of low-resource languages.
Most samples in the finetuning dataset also involve only one \pl.
Having multiple \pls in one data sample allows backpropagation of gradients across different \expertadapters, which can further improve knowledge sharing and specialization.
Code translation may be one source of such multilingual data, however, our preliminary experiments show that the quality of finetuning data is critical.
In the future, we will identify high-quality multilingual finetuning datasets to improve \Tool's performance.

The ideal usage scenario of \Tool is as an IDE/editor extension, where the extension is initialized with one copy of \basemodel, \sharedadapter, and \nladapter, and the user can download only the \expertadapters for the \pls that they plan to use (which are much smaller than downloading a language-specific finetuned model without using \Tool).  We plan to explore \Tool's ability to finetune newly added \expertadapters in future work.

\section{Related Work}
\label{sec:related}

\MyPara{LLMs for Multilingual Programming} Most LLMs for code in recent years~\cite{ChenETAL21Codex, CodeWhisperer, Copilot, luo2023wizardcoder, ZhengETAL23CodeGeeX, WeiETAL24Magicoder}, including the base model we used in our study---\deepseekcoder~\cite{GuoETAL24DeepSeekCoder}, are pretrained on code from multiple \pls, thus are multilingual by natural.
Researchers have been exploring applying LLMs to code translation~\cite{ZhangETAL23Coeditor,MacedoETAL25InterTrans,LachauxETAL20Transcoder}, a task involving two \pls.
Xu et al.~\cite{XuETAL22PolyCoder}'s PolyCoder was among the first to systematically evaluate the performance of multilingual LLMs for code.  However, all existing work uses the default transformer architecture, and ours is the first to propose an architecture that optimizes the efficiency of training multilingual LLMs for code.

To evaluate code LLMs' performance across multiple \pls, a number of multilingual benchmarks have been created, such as \humanevalpack~\cite{MuennighoffETAL24OctoPack} (which we used in our study), MultiPL-E~\cite{CassanoETAL23MultiPLE}, and HumanEval-X~\cite{ZhengETAL23CodeGeeX}.  They all come with a diverse set of \pls and executable test cases to evaluate the correctness of model outputs (following the practice of their origin dataset, HumanEval~\cite{ChenETAL21Codex}).  We adopted \humanevalpack in our study due to its multiple programming tasks and having consistent number of samples across all \pls.

\MyPara{Parameter-Efficient Finetuning of LLMs} Low-rank adaptation (\lora) and their variants~\cite{HuETAL21LoRA, kalajdzievski2023rankstabilizationscalingfactor, DettmersETAL23QLoRA, zhang2023lorafamemoryefficientlowrankadaptation,liu2024doraweightdecomposedlowrankadaptation, kopiczko2024vera,si2024floralowrankcorespace} are widely used for the efficient finetuning of LLMs by constraining trainable parameters in a low rank space.  In this work, we use \lora \adapters to hold language-agnostic and language-specific trainable parameters.  Researchers have explored improving the initialization strategy of \lora by using principle component analysis~\cite{MengETAL24Pissa, BalazyETAL24LoRAXS}, which we also adopted in \Tool to bootstrap the distribution of programming-language-related knowledge among the \adapters.

\MyPara{Mixture-of-Experts (MoE) Architecture} The MoE architecture scales up the number of parameters by using a set of expert modules that will be conditionally activated based on input tokens, and has been shown to be effective on many tasks including programming~\cite{xue2024openmoeearlyeffortopen, JiangETAL24Mixtral,muennighoff2024olmoeopenmixtureofexpertslanguage,fedus2022reviewsparseexpertmodels,fedus2022switchtransformersscalingtrillion,du2022glamefficientscalinglanguage,zhou2022mixtureofexpertsexpertchoicerouting,puigcerver2024sparsesoftmixturesexperts,zhou2024brainformerstradingsimplicityefficiency,zoph2022stmoedesigningstabletransferable}.
Recent work also proposed using \lora in MoE~\cite{HuangETAL24LoraHub,DouETAL24LoRAMoE,ZadouriETAL24MoLoRA,ZhaoETAL25MoSLD}, which motivated the design of \Tool.
Our \Tool architecture is in spirit similar to MoE, but with a few key differences: (1)~the expert modules in MoE are not pre-assigned to specific tasks and all experts must be present during inference, while each \expertadapter in \Tool is in charge of a specific \pl and only the relevant experts are needed during inference; (2)~MoE learns a ``router'' gating network to determine which expert to activate, while \Tool determines which \expertadapter to activate based on prior context or user specification (Section~\ref{sec:technique:inference}).
MoE architecture can also be initialized from an existing pretrained LLM, using a technique known as upcycling~\cite{HeETAL24Upcycling, JiangETAL24Mixtral}; a common limitation of upcycling is the resulting expert modules become very similar, which inspired the design of \sharedadapter in \Tool.

\section{Conclusion}
\label{sec:conclusion}

We proposed \Tool, the first model architecture designed for finetuning code LLMs for multiple \pls.
By using a combination of a \sharedadapter, \expertadapters, and an \nladapter, \Tool balances efficiency and specialization for multilingual programming.
The \adapters are initialized according to a principal-components-based strategy, leveraging the pretrained model's knowledge, and are jointly finetuned on a dataset spanning multiple \pls.
At inference, the appropriate \expertadapter is dynamically activated based on the token being processed.
We performed an extensive evaluation on a benchmark with three tasks: code \describe, \synthesis, and \translation.
We demonstrated that \Tool is more effective than the baselines of finetuning a single model shared by all \pls, or finetuning multiple language-specific models, and is even close to that of a fully finetuned model.
We also confirmed the effectiveness of our proposed architecture in disseminating knowledge across multiple programming languages.
We envision that \Tool will be more powerful when applied on larger models trained with more computational resources, which we will explore in the future.

\section*{Acknowledgments}
We thank Yu Liu and the anonymous reviewers for their comments and feedback.
This work is partially supported by the Natural Sciences and Engineering Research Council of Canada (NSERC) under funding reference number RGPIN-2024-04909 and the University of Waterloo start-up grant.

\bibliography{bib}

\end{document}

%% file: tables/numbers-exp.tex
\DefMacro{describe_mole-48-16_cpp}{22.76}
\DefMacro{describe_mole-48-16_go}{19.31}
\DefMacro{describe_mole-48-16_java}{31.30}
\DefMacro{describe_mole-48-16_js}{21.75}
\DefMacro{describe_mole-48-16_python}{33.54}
\DefMacro{describe_mole-48-16_rust}{13.82}
\DefMacro{synthesize_mole-48-16_cpp}{31.30}
\DefMacro{synthesize_mole-48-16_go}{27.24}
\DefMacro{synthesize_mole-48-16_java}{36.79}
\DefMacro{synthesize_mole-48-16_js}{36.59}
\DefMacro{synthesize_mole-48-16_python}{36.18}
\DefMacro{synthesize_mole-48-16_rust}{19.92}
\DefMacro{translation_mole-48-16_cpp_go}{61.79}
\DefMacro{translation_mole-48-16_cpp_java}{81.71}
\DefMacro{translation_mole-48-16_cpp_js}{74.19}
\DefMacro{translation_mole-48-16_cpp_python}{72.36}
\DefMacro{translation_mole-48-16_cpp_rust}{46.34}
\DefMacro{translation_mole-48-16_go_cpp}{57.52}
\DefMacro{translation_mole-48-16_go_java}{69.51}
\DefMacro{translation_mole-48-16_go_js}{69.92}
\DefMacro{translation_mole-48-16_go_python}{80.08}
\DefMacro{translation_mole-48-16_go_rust}{36.18}
\DefMacro{translation_mole-48-16_java_cpp}{63.21}
\DefMacro{translation_mole-48-16_java_go}{51.02}
\DefMacro{translation_mole-48-16_java_js}{77.64}
\DefMacro{translation_mole-48-16_java_python}{79.88}
\DefMacro{translation_mole-48-16_java_rust}{41.26}
\DefMacro{translation_mole-48-16_js_cpp}{64.02}
\DefMacro{translation_mole-48-16_js_go}{50.61}
\DefMacro{translation_mole-48-16_js_java}{71.14}
\DefMacro{translation_mole-48-16_js_python}{75.20}
\DefMacro{translation_mole-48-16_js_rust}{32.72}
\DefMacro{translation_mole-48-16_python_cpp}{51.22}
\DefMacro{translation_mole-48-16_python_go}{39.23}
\DefMacro{translation_mole-48-16_python_java}{62.60}
\DefMacro{translation_mole-48-16_python_js}{65.24}
\DefMacro{translation_mole-48-16_python_rust}{31.50}
\DefMacro{translation_mole-48-16_rust_cpp}{64.23}
\DefMacro{translation_mole-48-16_rust_go}{45.53}
\DefMacro{translation_mole-48-16_rust_java}{65.65}
\DefMacro{translation_mole-48-16_rust_js}{66.87}
\DefMacro{translation_mole-48-16_rust_python}{70.33}
\DefMacro{translation_mole-32-32_cpp_go}{61.38}
\DefMacro{translation_mole-32-32_cpp_java}{81.50}
\DefMacro{translation_mole-32-32_cpp_js}{76.02}
\DefMacro{translation_mole-32-32_cpp_python}{71.95}
\DefMacro{translation_mole-32-32_cpp_rust}{44.11}
\DefMacro{translation_mole-32-32_go_cpp}{56.91}
\DefMacro{translation_mole-32-32_go_java}{68.09}
\DefMacro{translation_mole-32-32_go_js}{70.12}
\DefMacro{translation_mole-32-32_go_python}{80.28}
\DefMacro{translation_mole-32-32_go_rust}{38.21}
\DefMacro{translation_mole-32-32_java_cpp}{63.21}
\DefMacro{translation_mole-32-32_java_go}{49.80}
\DefMacro{translation_mole-32-32_java_js}{80.08}
\DefMacro{translation_mole-32-32_java_python}{82.32}
\DefMacro{translation_mole-32-32_java_rust}{40.85}
\DefMacro{translation_mole-32-32_js_cpp}{61.38}
\DefMacro{translation_mole-32-32_js_go}{49.59}
\DefMacro{translation_mole-32-32_js_java}{70.93}
\DefMacro{translation_mole-32-32_js_python}{75.41}
\DefMacro{translation_mole-32-32_js_rust}{33.94}
\DefMacro{translation_mole-32-32_python_cpp}{50.41}
\DefMacro{translation_mole-32-32_python_go}{37.80}
\DefMacro{translation_mole-32-32_python_java}{62.60}
\DefMacro{translation_mole-32-32_python_js}{65.45}
\DefMacro{translation_mole-32-32_python_rust}{31.71}
\DefMacro{translation_mole-32-32_rust_cpp}{64.63}
\DefMacro{translation_mole-32-32_rust_go}{46.14}
\DefMacro{translation_mole-32-32_rust_java}{62.60}
\DefMacro{translation_mole-32-32_rust_js}{66.67}
\DefMacro{translation_mole-32-32_rust_python}{67.68}
\DefMacro{translation_full-ft_cpp_go}{62.20}
\DefMacro{translation_full-ft_cpp_java}{79.27}
\DefMacro{translation_full-ft_cpp_js}{73.58}
\DefMacro{translation_full-ft_cpp_python}{71.54}
\DefMacro{translation_full-ft_cpp_rust}{43.70}
\DefMacro{translation_full-ft_go_cpp}{57.52}
\DefMacro{translation_full-ft_go_java}{69.11}
\DefMacro{translation_full-ft_go_js}{72.15}
\DefMacro{translation_full-ft_go_python}{80.08}
\DefMacro{translation_full-ft_go_rust}{40.24}
\DefMacro{translation_full-ft_java_cpp}{65.65}
\DefMacro{translation_full-ft_java_go}{50.00}
\DefMacro{translation_full-ft_java_js}{74.19}
\DefMacro{translation_full-ft_java_python}{82.52}
\DefMacro{translation_full-ft_java_rust}{46.95}
\DefMacro{translation_full-ft_js_cpp}{62.80}
\DefMacro{translation_full-ft_js_go}{52.64}
\DefMacro{translation_full-ft_js_java}{69.51}
\DefMacro{translation_full-ft_js_python}{75.61}
\DefMacro{translation_full-ft_js_rust}{36.59}
\DefMacro{translation_full-ft_python_cpp}{49.19}
\DefMacro{translation_full-ft_python_go}{37.80}
\DefMacro{translation_full-ft_python_java}{59.15}
\DefMacro{translation_full-ft_python_js}{64.02}
\DefMacro{translation_full-ft_python_rust}{34.35}
\DefMacro{translation_full-ft_rust_cpp}{64.23}
\DefMacro{translation_full-ft_rust_go}{46.95}
\DefMacro{translation_full-ft_rust_java}{64.23}
\DefMacro{translation_full-ft_rust_js}{63.41}
\DefMacro{translation_full-ft_rust_python}{69.92}
\DefMacro{translation_all-lang_cpp_go}{59.96}
\DefMacro{translation_all-lang_cpp_java}{77.44}
\DefMacro{translation_all-lang_cpp_js}{75.00}
\DefMacro{translation_all-lang_cpp_python}{66.87}
\DefMacro{translation_all-lang_cpp_rust}{43.09}
\DefMacro{translation_all-lang_go_cpp}{56.50}
\DefMacro{translation_all-lang_go_java}{66.87}
\DefMacro{translation_all-lang_go_js}{69.31}
\DefMacro{translation_all-lang_go_python}{80.49}
\DefMacro{translation_all-lang_go_rust}{36.18}
\DefMacro{translation_all-lang_java_cpp}{65.04}
\DefMacro{translation_all-lang_java_go}{51.02}
\DefMacro{translation_all-lang_java_js}{75.20}
\DefMacro{translation_all-lang_java_python}{83.54}
\DefMacro{translation_all-lang_java_rust}{40.45}
\DefMacro{translation_all-lang_js_cpp}{63.01}
\DefMacro{translation_all-lang_js_go}{51.83}
\DefMacro{translation_all-lang_js_java}{72.97}
\DefMacro{translation_all-lang_js_python}{75.20}
\DefMacro{translation_all-lang_js_rust}{32.11}
\DefMacro{translation_all-lang_python_cpp}{48.58}
\DefMacro{translation_all-lang_python_go}{40.45}
\DefMacro{translation_all-lang_python_java}{64.63}
\DefMacro{translation_all-lang_python_js}{66.26}
\DefMacro{translation_all-lang_python_rust}{32.52}
\DefMacro{translation_all-lang_rust_cpp}{62.60}
\DefMacro{translation_all-lang_rust_go}{47.15}
\DefMacro{translation_all-lang_rust_java}{62.20}
\DefMacro{translation_all-lang_rust_js}{59.55}
\DefMacro{translation_all-lang_rust_python}{67.28}
\DefMacro{describe_mole-48-16-cpp_cpp}{22.76}
\DefMacro{describe_mole-48-16-cpp_go}{20.53}
\DefMacro{describe_mole-48-16-cpp_java}{30.89}
\DefMacro{describe_mole-48-16-cpp_js}{23.78}
\DefMacro{describe_mole-48-16-cpp_python}{30.89}
\DefMacro{describe_mole-48-16-cpp_rust}{12.40}
\DefMacro{synthesize_mole-48-16-cpp_cpp}{31.30}
\DefMacro{synthesize_mole-48-16-cpp_go}{27.44}
\DefMacro{synthesize_mole-48-16-cpp_java}{37.20}
\DefMacro{synthesize_mole-48-16-cpp_js}{35.57}
\DefMacro{synthesize_mole-48-16-cpp_python}{35.77}
\DefMacro{synthesize_mole-48-16-cpp_rust}{19.92}
\DefMacro{describe_mole-48-16-go_cpp}{22.97}
\DefMacro{describe_mole-48-16-go_go}{19.11}
\DefMacro{describe_mole-48-16-go_java}{30.49}
\DefMacro{describe_mole-48-16-go_js}{22.36}
\DefMacro{describe_mole-48-16-go_python}{33.74}
\DefMacro{describe_mole-48-16-go_rust}{14.63}
\DefMacro{synthesize_mole-48-16-go_cpp}{32.11}
\DefMacro{synthesize_mole-48-16-go_go}{27.44}
\DefMacro{synthesize_mole-48-16-go_java}{38.41}
\DefMacro{synthesize_mole-48-16-go_js}{34.76}
\DefMacro{synthesize_mole-48-16-go_python}{35.77}
\DefMacro{synthesize_mole-48-16-go_rust}{21.14}
\DefMacro{describe_mole-48-16-java_cpp}{21.95}
\DefMacro{describe_mole-48-16-java_go}{17.07}
\DefMacro{describe_mole-48-16-java_java}{31.30}
\DefMacro{describe_mole-48-16-java_js}{22.97}
\DefMacro{describe_mole-48-16-java_python}{31.71}
\DefMacro{describe_mole-48-16-java_rust}{12.80}
\DefMacro{synthesize_mole-48-16-java_cpp}{32.11}
\DefMacro{synthesize_mole-48-16-java_go}{26.42}
\DefMacro{synthesize_mole-48-16-java_java}{36.59}
\DefMacro{synthesize_mole-48-16-java_js}{35.37}
\DefMacro{synthesize_mole-48-16-java_python}{33.74}
\DefMacro{synthesize_mole-48-16-java_rust}{19.72}
\DefMacro{describe_mole-48-16-js_cpp}{22.97}
\DefMacro{describe_mole-48-16-js_go}{20.12}
\DefMacro{describe_mole-48-16-js_java}{30.28}
\DefMacro{describe_mole-48-16-js_js}{21.95}
\DefMacro{describe_mole-48-16-js_python}{32.93}
\DefMacro{describe_mole-48-16-js_rust}{14.02}
\DefMacro{synthesize_mole-48-16-js_cpp}{31.10}
\DefMacro{synthesize_mole-48-16-js_go}{27.85}
\DefMacro{synthesize_mole-48-16-js_java}{38.21}
\DefMacro{synthesize_mole-48-16-js_js}{36.79}
\DefMacro{synthesize_mole-48-16-js_python}{33.74}
\DefMacro{synthesize_mole-48-16-js_rust}{20.12}
\DefMacro{describe_mole-48-16-python_cpp}{25.00}
\DefMacro{describe_mole-48-16-python_go}{18.70}
\DefMacro{describe_mole-48-16-python_java}{29.47}
\DefMacro{describe_mole-48-16-python_js}{24.39}
\DefMacro{describe_mole-48-16-python_python}{33.54}
\DefMacro{describe_mole-48-16-python_rust}{13.82}
\DefMacro{synthesize_mole-48-16-python_cpp}{33.74}
\DefMacro{synthesize_mole-48-16-python_go}{27.24}
\DefMacro{synthesize_mole-48-16-python_java}{36.79}
\DefMacro{synthesize_mole-48-16-python_js}{35.77}
\DefMacro{synthesize_mole-48-16-python_python}{36.18}
\DefMacro{synthesize_mole-48-16-python_rust}{20.53}
\DefMacro{describe_mole-48-16-rust_cpp}{22.56}
\DefMacro{describe_mole-48-16-rust_go}{19.92}
\DefMacro{describe_mole-48-16-rust_java}{30.49}
\DefMacro{describe_mole-48-16-rust_js}{22.15}
\DefMacro{describe_mole-48-16-rust_python}{30.69}
\DefMacro{describe_mole-48-16-rust_rust}{13.82}
\DefMacro{synthesize_mole-48-16-rust_cpp}{32.52}
\DefMacro{synthesize_mole-48-16-rust_go}{27.44}
\DefMacro{synthesize_mole-48-16-rust_java}{38.01}
\DefMacro{synthesize_mole-48-16-rust_js}{32.93}
\DefMacro{synthesize_mole-48-16-rust_python}{35.16}
\DefMacro{synthesize_mole-48-16-rust_rust}{19.92}
\DefMacro{describe_mole-48-16-nl_cpp}{21.95}
\DefMacro{describe_mole-48-16-nl_go}{19.72}
\DefMacro{describe_mole-48-16-nl_java}{28.66}
\DefMacro{describe_mole-48-16-nl_js}{22.97}
\DefMacro{describe_mole-48-16-nl_python}{31.30}
\DefMacro{describe_mole-48-16-nl_rust}{11.38}
\DefMacro{synthesize_mole-48-16-nl_cpp}{33.13}
\DefMacro{synthesize_mole-48-16-nl_go}{26.42}
\DefMacro{synthesize_mole-48-16-nl_java}{40.24}
\DefMacro{synthesize_mole-48-16-nl_js}{36.38}
\DefMacro{synthesize_mole-48-16-nl_python}{36.59}
\DefMacro{synthesize_mole-48-16-nl_rust}{14.23}
\DefMacro{describe_mole-48-16-pl_cpp}{16.26}
\DefMacro{describe_mole-48-16-pl_go}{11.38}
\DefMacro{describe_mole-48-16-pl_java}{17.07}
\DefMacro{describe_mole-48-16-pl_js}{14.02}
\DefMacro{describe_mole-48-16-pl_python}{16.67}
\DefMacro{describe_mole-48-16-pl_rust}{10.77}
\DefMacro{synthesize_mole-48-16-pl_cpp}{26.42}
\DefMacro{synthesize_mole-48-16-pl_go}{14.63}
\DefMacro{synthesize_mole-48-16-pl_java}{29.07}
\DefMacro{synthesize_mole-48-16-pl_js}{26.02}
\DefMacro{synthesize_mole-48-16-pl_python}{24.39}
\DefMacro{synthesize_mole-48-16-pl_rust}{14.23}
\DefMacro{describe_mole-48-16-shared-last_cpp}{21.14}
\DefMacro{describe_mole-48-16-shared-last_go}{16.46}
\DefMacro{describe_mole-48-16-shared-last_java}{24.59}
\DefMacro{describe_mole-48-16-shared-last_js}{18.09}
\DefMacro{describe_mole-48-16-shared-last_python}{27.64}
\DefMacro{describe_mole-48-16-shared-last_rust}{14.23}
\DefMacro{synthesize_mole-48-16-shared-last_cpp}{31.50}
\DefMacro{synthesize_mole-48-16-shared-last_go}{25.81}
\DefMacro{synthesize_mole-48-16-shared-last_java}{37.20}
\DefMacro{synthesize_mole-48-16-shared-last_js}{35.98}
\DefMacro{synthesize_mole-48-16-shared-last_python}{34.55}
\DefMacro{synthesize_mole-48-16-shared-last_rust}{18.50}
\DefMacro{describe_mole-48-16-std-init_cpp}{23.37}
\DefMacro{describe_mole-48-16-std-init_go}{16.26}
\DefMacro{describe_mole-48-16-std-init_java}{28.86}
\DefMacro{describe_mole-48-16-std-init_js}{23.58}
\DefMacro{describe_mole-48-16-std-init_python}{32.72}
\DefMacro{describe_mole-48-16-std-init_rust}{14.84}
\DefMacro{synthesize_mole-48-16-std-init_cpp}{33.33}
\DefMacro{synthesize_mole-48-16-std-init_go}{25.81}
\DefMacro{synthesize_mole-48-16-std-init_java}{40.24}
\DefMacro{synthesize_mole-48-16-std-init_js}{37.40}
\DefMacro{synthesize_mole-48-16-std-init_python}{34.55}
\DefMacro{synthesize_mole-48-16-std-init_rust}{19.51}
\DefMacro{describe_mole-48-16-nl-expert_cpp}{21.34}
\DefMacro{describe_mole-48-16-nl-expert_go}{19.11}
\DefMacro{describe_mole-48-16-nl-expert_java}{30.08}
\DefMacro{describe_mole-48-16-nl-expert_js}{22.15}
\DefMacro{describe_mole-48-16-nl-expert_python}{30.49}
\DefMacro{describe_mole-48-16-nl-expert_rust}{13.21}
\DefMacro{synthesize_mole-48-16-nl-expert_cpp}{31.71}
\DefMacro{synthesize_mole-48-16-nl-expert_go}{26.42}
\DefMacro{synthesize_mole-48-16-nl-expert_java}{36.99}
\DefMacro{synthesize_mole-48-16-nl-expert_js}{33.94}
\DefMacro{synthesize_mole-48-16-nl-expert_python}{34.76}
\DefMacro{synthesize_mole-48-16-nl-expert_rust}{19.31}
\DefMacro{translation_mole-48-16-force-src_cpp_go}{60.98}
\DefMacro{translation_mole-48-16-force-src_cpp_java}{81.30}
\DefMacro{translation_mole-48-16-force-src_cpp_js}{73.37}
\DefMacro{translation_mole-48-16-force-src_cpp_python}{70.33}
\DefMacro{translation_mole-48-16-force-src_cpp_rust}{37.80}
\DefMacro{translation_mole-48-16-force-src_go_cpp}{56.71}
\DefMacro{translation_mole-48-16-force-src_go_java}{67.89}
\DefMacro{translation_mole-48-16-force-src_go_js}{68.09}
\DefMacro{translation_mole-48-16-force-src_go_python}{78.86}
\DefMacro{translation_mole-48-16-force-src_go_rust}{34.76}
\DefMacro{translation_mole-48-16-force-src_java_cpp}{62.80}
\DefMacro{translation_mole-48-16-force-src_java_go}{50.81}
\DefMacro{translation_mole-48-16-force-src_java_js}{76.42}
\DefMacro{translation_mole-48-16-force-src_java_python}{80.69}
\DefMacro{translation_mole-48-16-force-src_java_rust}{41.26}
\DefMacro{translation_mole-48-16-force-src_js_cpp}{65.45}
\DefMacro{translation_mole-48-16-force-src_js_go}{52.24}
\DefMacro{translation_mole-48-16-force-src_js_java}{68.09}
\DefMacro{translation_mole-48-16-force-src_js_python}{74.39}
\DefMacro{translation_mole-48-16-force-src_js_rust}{32.11}
\DefMacro{translation_mole-48-16-force-src_python_cpp}{48.78}
\DefMacro{translation_mole-48-16-force-src_python_go}{38.82}
\DefMacro{translation_mole-48-16-force-src_python_java}{65.04}
\DefMacro{translation_mole-48-16-force-src_python_js}{64.02}
\DefMacro{translation_mole-48-16-force-src_python_rust}{31.71}
\DefMacro{translation_mole-48-16-force-src_rust_cpp}{60.77}
\DefMacro{translation_mole-48-16-force-src_rust_go}{41.46}
\DefMacro{translation_mole-48-16-force-src_rust_java}{67.48}
\DefMacro{translation_mole-48-16-force-src_rust_js}{57.32}
\DefMacro{translation_mole-48-16-force-src_rust_python}{65.24}
\DefMacro{translation_mole-48-16-force-tgt_cpp_go}{63.01}
\DefMacro{translation_mole-48-16-force-tgt_cpp_java}{80.49}
\DefMacro{translation_mole-48-16-force-tgt_cpp_js}{75.00}
\DefMacro{translation_mole-48-16-force-tgt_cpp_python}{72.76}
\DefMacro{translation_mole-48-16-force-tgt_cpp_rust}{45.73}
\DefMacro{translation_mole-48-16-force-tgt_go_cpp}{56.91}
\DefMacro{translation_mole-48-16-force-tgt_go_java}{71.14}
\DefMacro{translation_mole-48-16-force-tgt_go_js}{69.72}
\DefMacro{translation_mole-48-16-force-tgt_go_python}{79.07}
\DefMacro{translation_mole-48-16-force-tgt_go_rust}{36.38}
\DefMacro{translation_mole-48-16-force-tgt_java_cpp}{64.43}
\DefMacro{translation_mole-48-16-force-tgt_java_go}{53.25}
\DefMacro{translation_mole-48-16-force-tgt_java_js}{77.64}
\DefMacro{translation_mole-48-16-force-tgt_java_python}{79.27}
\DefMacro{translation_mole-48-16-force-tgt_java_rust}{39.43}
\DefMacro{translation_mole-48-16-force-tgt_js_cpp}{64.02}
\DefMacro{translation_mole-48-16-force-tgt_js_go}{52.85}
\DefMacro{translation_mole-48-16-force-tgt_js_java}{72.15}
\DefMacro{translation_mole-48-16-force-tgt_js_python}{75.61}
\DefMacro{translation_mole-48-16-force-tgt_js_rust}{31.50}
\DefMacro{translation_mole-48-16-force-tgt_python_cpp}{52.85}
\DefMacro{translation_mole-48-16-force-tgt_python_go}{39.84}
\DefMacro{translation_mole-48-16-force-tgt_python_java}{63.41}
\DefMacro{translation_mole-48-16-force-tgt_python_js}{65.04}
\DefMacro{translation_mole-48-16-force-tgt_python_rust}{31.10}
\DefMacro{translation_mole-48-16-force-tgt_rust_cpp}{64.43}
\DefMacro{translation_mole-48-16-force-tgt_rust_go}{45.53}
\DefMacro{translation_mole-48-16-force-tgt_rust_java}{64.84}
\DefMacro{translation_mole-48-16-force-tgt_rust_js}{64.23}
\DefMacro{translation_mole-48-16-force-tgt_rust_python}{68.09}

%% file: tables/numbers-exp-manual.tex
\DefMacro{describe_pretrained_cpp}{16.46}
\DefMacro{describe_pretrained_go}{8.54}
\DefMacro{describe_pretrained_java}{29.88}
\DefMacro{describe_pretrained_js}{15.24}
\DefMacro{describe_pretrained_python}{19.51}
\DefMacro{describe_pretrained_rust}{10.37}

\DefMacro{synthesize_pretrained_cpp}{33.54}
\DefMacro{synthesize_pretrained_go}{26.22}
\DefMacro{synthesize_pretrained_java}{31.71}
\DefMacro{synthesize_pretrained_js}{32.32}
\DefMacro{synthesize_pretrained_python}{33.54}
\DefMacro{synthesize_pretrained_rust}{17.07}

\DefMacro{translation_pretrained_cpp_go}{42.68}
\DefMacro{translation_pretrained_cpp_java}{63.41}
\DefMacro{translation_pretrained_cpp_js}{65.85}
\DefMacro{translation_pretrained_cpp_python}{52.44}
\DefMacro{translation_pretrained_cpp_rust}{31.71}
\DefMacro{translation_pretrained_go_cpp}{52.44}
\DefMacro{translation_pretrained_go_java}{62.20}
\DefMacro{translation_pretrained_go_js}{62.20}
\DefMacro{translation_pretrained_go_python}{62.80}
\DefMacro{translation_pretrained_go_rust}{27.44}
\DefMacro{translation_pretrained_java_cpp}{59.76}
\DefMacro{translation_pretrained_java_go}{45.73}
\DefMacro{translation_pretrained_java_js}{67.07}
\DefMacro{translation_pretrained_java_python}{73.17}
\DefMacro{translation_pretrained_java_rust}{35.37}
\DefMacro{translation_pretrained_js_cpp}{53.66}
\DefMacro{translation_pretrained_js_go}{40.85}
\DefMacro{translation_pretrained_js_java}{48.78}
\DefMacro{translation_pretrained_js_python}{57.32}
\DefMacro{translation_pretrained_js_rust}{28.66}
\DefMacro{translation_pretrained_python_cpp}{45.73}
\DefMacro{translation_pretrained_python_go}{26.83}
\DefMacro{translation_pretrained_python_java}{59.15}
\DefMacro{translation_pretrained_python_js}{53.05}
\DefMacro{translation_pretrained_python_rust}{29.88}
\DefMacro{translation_pretrained_rust_cpp}{57.32}
\DefMacro{translation_pretrained_rust_go}{40.24}
\DefMacro{translation_pretrained_rust_java}{57.93}
\DefMacro{translation_pretrained_rust_js}{16.46}
\DefMacro{translation_pretrained_rust_python}{45.12}

\DefMacro{describe_per-lang_cpp}{20.53}
\DefMacro{describe_per-lang_go}{16.46}
\DefMacro{describe_per-lang_java}{30.69}
\DefMacro{describe_per-lang_js}{22.76}
\DefMacro{describe_per-lang_python}{33.54}
\DefMacro{describe_per-lang_rust}{11.59}

\DefMacro{synthesize_per-lang_cpp}{32.93}
\DefMacro{synthesize_per-lang_go}{24.80}
\DefMacro{synthesize_per-lang_java}{38.82}
\DefMacro{synthesize_per-lang_js}{33.13}
\DefMacro{synthesize_per-lang_python}{34.76}
\DefMacro{synthesize_per-lang_rust}{21.34}

\DefMacro{describe_all-lang_cpp}{23.17}
\DefMacro{describe_all-lang_go}{19.11}
\DefMacro{describe_all-lang_java}{29.27}
\DefMacro{describe_all-lang_js}{20.12}
\DefMacro{describe_all-lang_python}{31.50}
\DefMacro{describe_all-lang_rust}{11.38}

\DefMacro{synthesize_all-lang_cpp}{30.89}
\DefMacro{synthesize_all-lang_go}{27.03}
\DefMacro{synthesize_all-lang_java}{36.18}
\DefMacro{synthesize_all-lang_js}{36.59}
\DefMacro{synthesize_all-lang_python}{35.37}
\DefMacro{synthesize_all-lang_rust}{17.68}

\DefMacro{describe_mole-64-0_cpp}{22.36}
\DefMacro{describe_mole-64-0_go}{18.09}
\DefMacro{describe_mole-64-0_java}{27.64}
\DefMacro{describe_mole-64-0_js}{22.97}
\DefMacro{describe_mole-64-0_python}{31.50}
\DefMacro{describe_mole-64-0_rust}{12.40}

\DefMacro{synthesize_mole-64-0_cpp}{31.50}
\DefMacro{synthesize_mole-64-0_go}{25.41}
\DefMacro{synthesize_mole-64-0_java}{35.98}
\DefMacro{synthesize_mole-64-0_js}{34.15}
\DefMacro{synthesize_mole-64-0_python}{34.96}
\DefMacro{synthesize_mole-64-0_rust}{18.29}

\DefMacro{describe_mole-56-8_cpp}{22.97}
\DefMacro{describe_mole-56-8_go}{20.73}
\DefMacro{describe_mole-56-8_java}{29.07}
\DefMacro{describe_mole-56-8_js}{22.56}
\DefMacro{describe_mole-56-8_python}{31.91}
\DefMacro{describe_mole-56-8_rust}{11.99}

\DefMacro{synthesize_mole-56-8_cpp}{32.11}
\DefMacro{synthesize_mole-56-8_go}{27.24}
\DefMacro{synthesize_mole-56-8_java}{35.16}
\DefMacro{synthesize_mole-56-8_js}{34.96}
\DefMacro{synthesize_mole-56-8_python}{35.77}
\DefMacro{synthesize_mole-56-8_rust}{20.73}

\DefMacro{describe_mole-32-32_cpp}{20.93}
\DefMacro{describe_mole-32-32_go}{21.14}
\DefMacro{describe_mole-32-32_java}{30.69}
\DefMacro{describe_mole-32-32_js}{24.19}
\DefMacro{describe_mole-32-32_python}{31.50}
\DefMacro{describe_mole-32-32_rust}{13.01}

\DefMacro{synthesize_mole-32-32_cpp}{30.49}
\DefMacro{synthesize_mole-32-32_go}{26.63}
\DefMacro{synthesize_mole-32-32_java}{38.21}
\DefMacro{synthesize_mole-32-32_js}{38.21}
\DefMacro{synthesize_mole-32-32_python}{35.98}
\DefMacro{synthesize_mole-32-32_rust}{20.33}

\DefMacro{describe_full-ft_cpp}{22.76}
\DefMacro{describe_full-ft_go}{20.33}
\DefMacro{describe_full-ft_java}{27.03}
\DefMacro{describe_full-ft_js}{26.42}
\DefMacro{describe_full-ft_python}{31.50}
\DefMacro{describe_full-ft_rust}{14.23}

\DefMacro{synthesize_full-ft_cpp}{35.37}
\DefMacro{synthesize_full-ft_go}{28.86}
\DefMacro{synthesize_full-ft_java}{37.80}
\DefMacro{synthesize_full-ft_js}{39.43}
\DefMacro{synthesize_full-ft_python}{39.84}
\DefMacro{synthesize_full-ft_rust}{22.56}

%% file: tables/numbers-exp-computed.tex
\DefMacro{describe_pretrained_AVG}{16.67}
\DefMacro{synthesize_pretrained_AVG}{29.07}
\DefMacro{describe_all-lang_AVG}{22.43}
\DefMacro{synthesize_all-lang_AVG}{30.62}
\DefMacro{describe_per-lang_AVG}{22.60}
\DefMacro{synthesize_per-lang_AVG}{30.96}
\DefMacro{describe_full-ft_AVG}{23.71}
\DefMacro{synthesize_full-ft_AVG}{33.98}
\DefMacro{describe_mole-32-32_AVG}{23.58}
\DefMacro{synthesize_mole-32-32_AVG}{31.64}
\DefMacro{describe_mole-48-16_AVG}{23.75}
\DefMacro{synthesize_mole-48-16_AVG}{31.34}
\DefMacro{describe_mole-56-8_AVG}{23.21}
\DefMacro{synthesize_mole-56-8_AVG}{31.00}
\DefMacro{describe_mole-64-0_AVG}{22.49}
\DefMacro{synthesize_mole-64-0_AVG}{30.05}
\DefMacro{describe_mole-48-16-cpp_AVG}{23.54}
\DefMacro{synthesize_mole-48-16-cpp_AVG}{31.20}
\DefMacro{describe_mole-48-16-go_AVG}{23.88}
\DefMacro{synthesize_mole-48-16-go_AVG}{31.61}
\DefMacro{describe_mole-48-16-java_AVG}{22.97}
\DefMacro{synthesize_mole-48-16-java_AVG}{30.66}
\DefMacro{describe_mole-48-16-js_AVG}{23.71}
\DefMacro{synthesize_mole-48-16-js_AVG}{31.30}
\DefMacro{describe_mole-48-16-python_AVG}{24.15}
\DefMacro{synthesize_mole-48-16-python_AVG}{31.71}
\DefMacro{describe_mole-48-16-rust_AVG}{23.27}
\DefMacro{synthesize_mole-48-16-rust_AVG}{31.00}
\DefMacro{describe_mole-48-16-nl_AVG}{22.66}
\DefMacro{synthesize_mole-48-16-nl_AVG}{31.17}
\DefMacro{describe_mole-48-16-pl_AVG}{14.36}
\DefMacro{synthesize_mole-48-16-pl_AVG}{22.46}
\DefMacro{describe_mole-48-16-shared-last_AVG}{20.36}
\DefMacro{synthesize_mole-48-16-shared-last_AVG}{30.59}
\DefMacro{describe_mole-48-16-std-init_AVG}{23.27}
\DefMacro{synthesize_mole-48-16-std-init_AVG}{31.81}
\DefMacro{describe_mole-48-16-nl-expert_AVG}{22.73}
\DefMacro{synthesize_mole-48-16-nl-expert_AVG}{30.52}
\DefMacro{translation_pretrained_cpp_AVG}{51.22}
\DefMacro{translation_pretrained_AVG_cpp}{53.78}
\DefMacro{translation_pretrained_go_AVG}{53.42}
\DefMacro{translation_pretrained_AVG_go}{39.27}
\DefMacro{translation_pretrained_java_AVG}{56.22}
\DefMacro{translation_pretrained_AVG_java}{58.29}
\DefMacro{translation_pretrained_js_AVG}{45.85}
\DefMacro{translation_pretrained_AVG_js}{52.93}
\DefMacro{translation_pretrained_python_AVG}{42.93}
\DefMacro{translation_pretrained_AVG_python}{58.17}
\DefMacro{translation_pretrained_rust_AVG}{43.41}
\DefMacro{translation_pretrained_AVG_rust}{30.61}
\DefMacro{translation_pretrained_AVG}{48.84}
\DefMacro{translation_all-lang_cpp_AVG}{64.47}
\DefMacro{translation_all-lang_AVG_cpp}{59.15}
\DefMacro{translation_all-lang_go_AVG}{61.87}
\DefMacro{translation_all-lang_AVG_go}{50.08}
\DefMacro{translation_all-lang_java_AVG}{63.05}
\DefMacro{translation_all-lang_AVG_java}{68.82}
\DefMacro{translation_all-lang_js_AVG}{59.02}
\DefMacro{translation_all-lang_AVG_js}{69.06}
\DefMacro{translation_all-lang_python_AVG}{50.49}
\DefMacro{translation_all-lang_AVG_python}{74.68}
\DefMacro{translation_all-lang_rust_AVG}{59.76}
\DefMacro{translation_all-lang_AVG_rust}{36.87}
\DefMacro{translation_all-lang_AVG}{59.78}
\DefMacro{translation_full-ft_cpp_AVG}{66.06}
\DefMacro{translation_full-ft_AVG_cpp}{59.88}
\DefMacro{translation_full-ft_go_AVG}{63.82}
\DefMacro{translation_full-ft_AVG_go}{49.92}
\DefMacro{translation_full-ft_java_AVG}{63.86}
\DefMacro{translation_full-ft_AVG_java}{68.25}
\DefMacro{translation_full-ft_js_AVG}{59.43}
\DefMacro{translation_full-ft_AVG_js}{69.47}
\DefMacro{translation_full-ft_python_AVG}{48.90}
\DefMacro{translation_full-ft_AVG_python}{75.93}
\DefMacro{translation_full-ft_rust_AVG}{61.75}
\DefMacro{translation_full-ft_AVG_rust}{40.37}
\DefMacro{translation_full-ft_AVG}{60.64}
\DefMacro{translation_mole-32-32_cpp_AVG}{66.99}
\DefMacro{translation_mole-32-32_AVG_cpp}{59.31}
\DefMacro{translation_mole-32-32_go_AVG}{62.72}
\DefMacro{translation_mole-32-32_AVG_go}{48.94}
\DefMacro{translation_mole-32-32_java_AVG}{63.25}
\DefMacro{translation_mole-32-32_AVG_java}{69.14}
\DefMacro{translation_mole-32-32_js_AVG}{58.25}
\DefMacro{translation_mole-32-32_AVG_js}{71.67}
\DefMacro{translation_mole-32-32_python_AVG}{49.59}
\DefMacro{translation_mole-32-32_AVG_python}{75.53}
\DefMacro{translation_mole-32-32_rust_AVG}{61.54}
\DefMacro{translation_mole-32-32_AVG_rust}{37.76}
\DefMacro{translation_mole-32-32_AVG}{60.39}
\DefMacro{translation_mole-48-16_cpp_AVG}{67.28}
\DefMacro{translation_mole-48-16_AVG_cpp}{60.04}
\DefMacro{translation_mole-48-16_go_AVG}{62.64}
\DefMacro{translation_mole-48-16_AVG_go}{49.64}
\DefMacro{translation_mole-48-16_java_AVG}{62.60}
\DefMacro{translation_mole-48-16_AVG_java}{70.12}
\DefMacro{translation_mole-48-16_js_AVG}{58.74}
\DefMacro{translation_mole-48-16_AVG_js}{70.77}
\DefMacro{translation_mole-48-16_python_AVG}{49.96}
\DefMacro{translation_mole-48-16_AVG_python}{75.57}
\DefMacro{translation_mole-48-16_rust_AVG}{62.52}
\DefMacro{translation_mole-48-16_AVG_rust}{37.60}
\DefMacro{translation_mole-48-16_AVG}{60.62}
\DefMacro{translation_mole-48-16-force-src_cpp_AVG}{64.76}
\DefMacro{translation_mole-48-16-force-src_AVG_cpp}{58.90}
\DefMacro{translation_mole-48-16-force-src_go_AVG}{61.26}
\DefMacro{translation_mole-48-16-force-src_AVG_go}{48.86}
\DefMacro{translation_mole-48-16-force-src_java_AVG}{62.40}
\DefMacro{translation_mole-48-16-force-src_AVG_java}{69.96}
\DefMacro{translation_mole-48-16-force-src_js_AVG}{58.46}
\DefMacro{translation_mole-48-16-force-src_AVG_js}{67.84}
\DefMacro{translation_mole-48-16-force-src_python_AVG}{49.67}
\DefMacro{translation_mole-48-16-force-src_AVG_python}{73.90}
\DefMacro{translation_mole-48-16-force-src_rust_AVG}{58.45}
\DefMacro{translation_mole-48-16-force-src_AVG_rust}{35.53}
\DefMacro{translation_mole-48-16-force-src_AVG}{59.17}
\DefMacro{translation_mole-48-16-force-tgt_cpp_AVG}{67.40}
\DefMacro{translation_mole-48-16-force-tgt_AVG_cpp}{60.53}
\DefMacro{translation_mole-48-16-force-tgt_go_AVG}{62.64}
\DefMacro{translation_mole-48-16-force-tgt_AVG_go}{50.90}
\DefMacro{translation_mole-48-16-force-tgt_java_AVG}{62.80}
\DefMacro{translation_mole-48-16-force-tgt_AVG_java}{70.41}
\DefMacro{translation_mole-48-16-force-tgt_js_AVG}{59.23}
\DefMacro{translation_mole-48-16-force-tgt_AVG_js}{70.33}
\DefMacro{translation_mole-48-16-force-tgt_python_AVG}{50.45}
\DefMacro{translation_mole-48-16-force-tgt_AVG_python}{74.96}
\DefMacro{translation_mole-48-16-force-tgt_rust_AVG}{61.42}
\DefMacro{translation_mole-48-16-force-tgt_AVG_rust}{36.83}
\DefMacro{translation_mole-48-16-force-tgt_AVG}{60.66}